# Machine Learning for Analyzing Atomic Force Microscopy (AFM) Images Generated from Polymer Blends

Aanish Paruchuri [1], Yunfei Wang[2], Xiaodan Gu[2], Arthi Jayaraman [3,4,5] *

1. Master of Science in Data Science Program, University of Delaware, Newark DE 19713
2. School of Polymer Science and Engineering, 118 College Drive, #5050 University of Southern Mississippi, Hattiesburg, MS 39406
3. Department of Chemical and Biomolecular Engineering, 150 Academy St, University of Delaware, Newark DE 19713
4. Department of Materials Science and Engineering, University of Delaware, Newark DE 19713
5. Data Science Institute, University of Delaware, Newark DE, 19713

* Corresponding authors **arthij@udel.edu**

**Abstract**

In this paper we present a new machine learning (ML) workflow with unsupervised learning techniques to identify domains within atomic force microscopy (AFM) images obtained from polymer films. The goal of the workflow is to identify the spatial location of the two types of polymer domains with little to no manual intervention (Task 1) and calculate the domain size distributions which in turn can help qualify the phase separated state of the material as macrophase or microphase ordered/disordered domains (Task 2). We briefly review existing approaches used in other fields - computer vision and signal processing – that can be applicable for the above tasks that happen frequently in the field of polymer science and engineering. We then test these approaches from computer vision and signal processing on the AFM image dataset to identify the strengths and limitations of each of these approaches for our first task. For our first domain segmentation task, we found that the workflow using discrete Fourier transform (DFT) or discrete cosine transform (DCT) with variance statistics as the feature works the best. The popular ResNet50 deep learning approach from computer vision field exhibited relatively poorer performance in the domain segmentation task for our AFM images as compared to the DFT and DCT based workflows. For the second task, for each of 144 input AFM images, we then used an existing *porespy* python package to calculate the domain size distribution from the output of that image from DFT-based workflow. The information and open-source codes we share in this paper can serve as a guide for researchers in the polymer and soft materials fields who need ML modeling and workflows for automated analyses of AFM images from polymer samples that may have crystalline/amorphous domains, sharp/rough interfaces between domains, or micro or macro-phase separated domains.

**I. Introduction**

Researchers working with macromolecular materials (e.g., block copolymers [1-6], polymer blends [7, 8], polymer nanocomposites [9-21]) rely on various characterization techniques to understand the multiscale structural arrangements these soft materials exhibit for various designs (polymer chemistry, architecture, molecular mass) and processing conditions (thermal annealing temperature, processing techniques, solvents). Unlike crystalline inorganic materials or precisely structured proteins, most synthetic polymers and soft materials exhibit a rich diversity of ordered and disordered structure(s) at various length scales, and in many cases with dispersity in the structural dimensions. The hierarchy and the distribution of structural dimensions together dictate the performance and effectiveness of the material in its eventual application or function. To gain an understanding of the structural hierarchy, polymer researchers often turn towards one or more microscopy techniques such as scanning electron microscopy (SEM), transmission electron microscopy (TEM), scanning transmission electron microscopy (STEM), cryo-TEM, and atomic force microscopy (AFM), together with other scattering and spectroscopy tools. The type of data one obtains from these microscopy techniques are typically two-dimensional (2D) images that convey the intended physical information about the structure of the sample being probed (e.g., chemical differences between various regions or domains, shapes and sizes of various domains, orientation of microdomains, softness/hardness of the regions, and physical roughness in the form of height maps). Traditionally, these images from microscopy measurements are manually interpreted, aided by, in most cases, proprietary software packages that are linked to the microscopy instrument. Such manual analyses and interpretation are subject to human biases, errors, and subjectivity; one can expect the errors and biases to grow with increasing sizes of the datasets and less time spent on analyzing each image. With the recent shift towards high-throughput experimentation and characterization and open science, there is a critical need to shift away from manual interpretation of the images or manual intervention during computational interpretation with proprietary software. Instead, there is a strong justification to move towards fast and objective automated open-source computational and machine learning (ML) workflows. Despite the many successes of ML workflows for analyses of structural characterization in inorganic and small-molecule organic materials fields [22-36], the analogous development and use of ML approaches customized for polymers and soft materials [37-39] is still relatively less prevalent. In this paper we present one such new ML-based workflow for objective interpretation of AFM images from polymer films, with minimum manual intervention.

For readers less familiar with AFM techniques, we briefly describe AFM technique's application in the context of soft materials and polymers. AFM is used to map the surface topography as well as mechanical response (phase image) of soft materials by measuring the spatial variations in interactions between the instrument tip and the surface (see perspective article Ref. [40]). AFM is an effective tool for detecting nanometer-scale 2D morphology of self-assembled polymers which is essential for designing materials for specific applications such as high-resolution etch masks, microelectronics, optics, and solar cells. (see for example papers on self-assembly of block copolymers in reviews [41, 42]) AFM is also utilized to measure the thickness and roughness of polymer materials by constructing three-dimensional (3D) mappings. [43, 44] However, the complexity of surface morphologies—such as diversity of morphologies, dispersity of length scales in polymer phase separation, presence of defects, and noise from the instrument makes manual analysis of AFM images challenging, subjective, and sometimes inaccurate. Consequently, there is a critical need for ML workflows that can automate the analysis of AFM characterization of polymer films.

In general, ML workflows for polymers/soft materials' characterization results can be classified as being 'predictive' or 'generative' in nature. While 'predictive' ML models project materials properties or classify materials based on their structural characterization, 'generative' ML models create synthetic forms of the characterization data; both of these enable downstream discovery of new materials with desired properties. There are many noteworthy studies showing predictive ML models used on AFM images for detecting or classifying features of interest [45-48], detecting defects [49], classifying structural information [50], and understanding morphology [51]. There are also studies showing use of generative ML models for increasing resolution [52, 53] and denoising of AFM data [54]. Generative models have also been used to convert one form of experimental characterization to another form [55-57] to address limited/disparate access to instrumentation resources and differences in interpretability of the two forms of characterization. Despite the promising solutions enabled by ML models, one major challenge that exists in the training of ML models is the need for manual labeling for supervised ML models. Supervised ML models leverage (manually) labeled data to learn patterns and relationships in the dataset. Manual labeling of experimental data for training the ML model is a time-consuming process and not always objective or error-free. To overcome the need for manual labeling, researchers have developed self-supervised learning techniques (e.g., see reference [58] for how self-supervised learning can be used in microscopy image analysis). Self-supervised learning techniques use unlabeled data to learn the pretext of the task without requiring extensive manually labeled data; however, this process requires extensive data and some steps in training still require manual input. In contrast to supervised and self-supervised ML models, in unsupervised ML approaches [59, 60], the model learns the patterns in the data by identifying relationships and organizing data into meaningful groups without requiring labeled data. The main challenge one would face with unsupervised learning is to formulate the problem in an efficient manner that would enable unsupervised ML algorithms to effectively uncover hidden patterns and structures within the data. Unlike supervised learning, where the task is clearly defined by labeled examples, unsupervised ML often requires careful engineering of how to represent the data.

Another challenge for training ML models for microscopy image analysis in the field of polymer science and engineering is insufficient experimental data for training the model. In many cases, it is not viable to have large experimental datasets due to limited availability and access to the instrument or insufficient time or limited material availability for sample preparation for the measurement. In such situations, one way to generate additional relevant data for training the ML models is through the use of synthetic (i.e., simulated) data that has features like a typical experimental measurement data (see for example reference [57] where such simulated data was used to train an ML model). One can also use an augmented data set combining simulated and experimentally measured data to train ML models. We note a particular benefit of most unsupervised ML methods that the dataset size is less of a constraint as compared to supervised ML models, making unsupervised ML models more suitable for our task at hand.

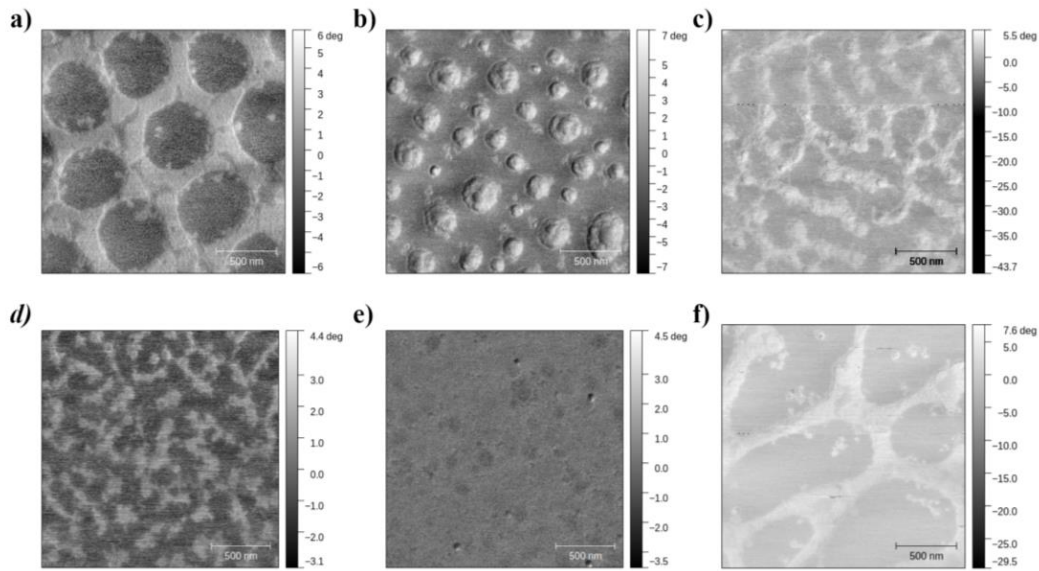

*Figure 1*: Examples form the dataset of AFM images of thin films of POEGMA-sb-PS spin-cast onto silicon wafers. Details of the sample preparation and AFM imaging are presented in **section III**.

In this paper we develop an ML workflow with unsupervised learning techniques to identify domains within this AFM image dataset obtained from films of supramolecular block copolymers. As supramolecular block polymers are formed from association of two types of homopolymers, we expect to see morphologies varying from large macro-phase separated (large) domains to ordered/disordered bicontinuous microphase separated (smaller) domains. (**Figure 1**) Through the developed ML workflow with unsupervised learning techniques we successfully identify the light and dark polymer domains in the input AFM images, and then quantify domain size distributions with little to no manual intervention. **Figure 2** describes the entire workflow.

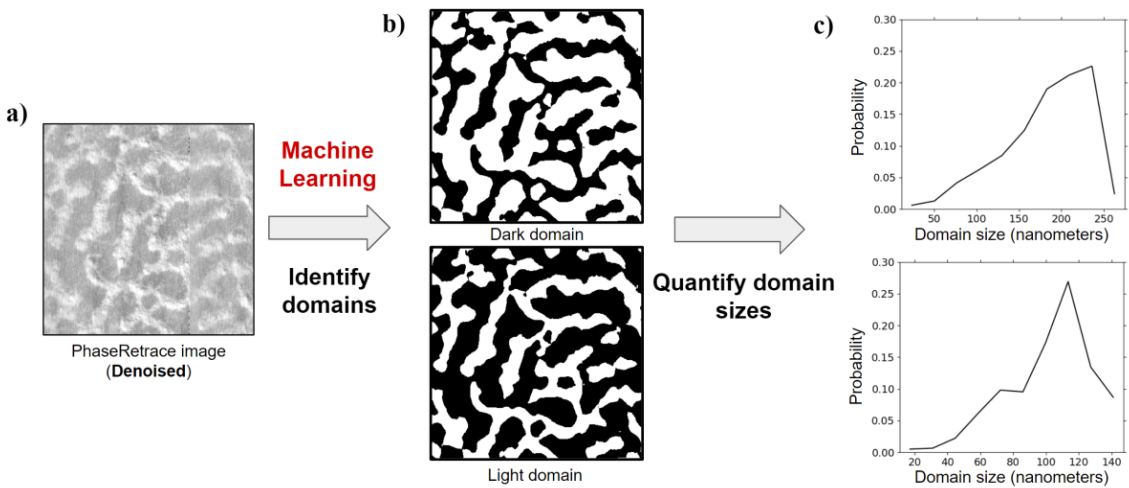

*Figure 2: Workflow for analyzing AFM images. We first apply machine learning techniques to denoised AFM phase images (part a is an example of a PhaseRetrace denoised image) and identify the positions of light and dark domains. The output is then two binary images, one for dark domains (top of part b) and one for light domains (bottom of part b). After the domains are identified, we use existing computational methods for domain size quantification to calculate the domain size distribution for light and dark domains (part c).*

Prior to showing the reader our developed workflow for identifying domains and quantifying the domain sizes, we first review (briefly) existing approaches in other fields - computer vision and signal processing – that are applicable for our task which researchers in polymer field frequent need to do during material characterization. We test many of these computer vision and signal processing approaches for our intended AFM image analysis tasks and through that process identify the strengths and limitations of each of these approaches for various AFM image analysis tasks for polymer samples. Even though we demonstrate the workflow only on the AFM images from polymer films comprised of blends of two associating homopolymers, we believe the information and open-source code we present in this paper should serve as a guide to polymer researchers who need ML modeling for analyses of AFM images from other polymer samples that may have crystalline/amorphous domains, sharp/rough interfaces between domains, or micro or macro- phase separated domains, as long as there is height or phase contrast.

**II. Computational Methods for AFM Image Analyses**

AFM images consist of a grid of data points that represent height or phase value obtained by the probe tip as it scans across the sample surface. This grid of data points in AFM images is similar to a grid of pixels in any everyday photo/image we capture, where each pixel in the image holds the light intensity values of the object or scene that is photographed. Essentially both the grids represent visually perceivable features and characteristics of the scanned surface or scene. Consequently, techniques used for image segmentation (defined as outlining object boundaries) and feature extraction in RGB (red, green, and blue) and grayscale images [45, 46, 50, 52] can be extended to AFM images. These techniques can elucidate surface texture which in turn could be used to distinguish materials' composition and structural arrangement at certain length scales. Texture refers to the visual patterns and structures (i.e., roughness, smoothness, and regularity in patterns). Texture analysis tasks involve quantifying patterns and extracting features to identify regions in an image. Various computational methods have been developed for texture analysis, including statistics-based [61, 62], transform-based[63-65], and more recently convolutional neural networks (CNN) [66, 67]. We want to understand some of these methods' strengths and limitations before we test their applicability to our specific task of AFM image analysis for polymer films' domain sizes and shapes (**Figure 1** and **Figure 2**).

The first category of methods includes wavelet, [68] Fourier, [69] and Radon [70] transforms which the field of signal processing relies on heavily. Typically, transforms are applied to convert the information from the input domain to a more favorable domain for easier analysis. The relevance of transforms to analysis of microscopy images arises from considering images from any microscopy measurement as a two-dimensional signal that can then be transformed into other easier-to-analyze forms. We describe the strengths and limitations of discrete Fourier transform, discrete cosine transform, two variants of discrete wavelet transform, and Radon transform in **section II.A**.

The second category of models and methods for automated image analysis comes from the field of computer vision. For automated image analysis, various ML models have been used to successfully identify and understand objects and people in stationary images and movies/videos.[71] For example, deep neural networks (DNN) including U-net[67], FastFCN[72] and Deeplab[73] are used for segmentation tasks in images. These models perform well on everyday images (of dogs, cats, furniture, etc.), but perform poorly when applied to field-specific (e.g., materials science, biomedical diagnostic) microscopy images. As these models are not trained on field-specific data, which for our paper is polymer film AFM images, the application of such models *as is* to analyze these AFM images is not an option. One can take DNN models trained on generic everyday objects and re-train a few layers of the neural network model to learn relevant features of the field-specific images; this is called *transfer learning*.[74] While transfer learning has been successful in many applications[75, 76], in most cases of transfer learning the re-training is done in a supervised manner with manual labeling, which is subjective and time-consuming. To be able to make an ML model plug-and-play for experimentalists, it is important to keep the nature of the solution *unsupervised*. Before we present the details of and results from unsupervised ML approaches that perform well for our specific task of AFM image analysis for polymer films' domain sizes and shapes, we present a brief review of the deep learning and cluster analyses both of which are relevant for our AFM image analyses tasks, in **section II.B**. and **II.C**, respectively. We also present the key parameters of these models and how to tune them to get desired analysis results.

**A. Domain transforms**

*Table 1: Definitions of common terms used in the context of domain transforms.*

| Term | Definition |
| --- | --- |
| Pixel | Smallest unit of image (single point in a grid) which holds intensity values. |
| Tile | Subsection of image on which domain transforms or DNN methods are applied to extract features. |
| Win factor | A factor that controls the size of tiles. Its defined as the ratio of input image size to tile size. Values of win factor lie in the range (0,1) where the tile size increases with increase in win factor. |

| Term | Definition |
| --- | --- |
| Texture | Visual patterns in an image. |
| Feature | Characteristics of data that are used to represent pattern or structure in images. |
| Padding | A practice of adding extra layers around the image to preserve spatial dimensions. |
| RGB | A digital image that represents color using red, green, and blue channels, where each pixel holds values of these three channels. |
| Wavelet | A mathematical function used to decompose data into its frequency components. |
| Natural image | A photograph or digital image that captures objects from the real world (e.g. car, furniture, animals etc.). |
| Domain transform | Operations that convert data from one representation to another. |
| K-means | Unsupervised ML method that clusters data based on similarity. |
| Index map | 2D matrix that signifies spatial position of domains. |
| Domains | Space or representation of data in which it is analyzed (e.g. time, frequency, spatial etc.). |
| Spatial domain | Representation of data where values are organized based on their spatial position. |
| Segmentation | A task of partitioning data in groups of distinct regions. |

In image analysis, domain transforms play a pivotal role in facilitating the conversion of images from spatial domains (in real space) to another domain (e.g., frequency space) which offers a more powerful means of extracting meaningful information pertaining to regions of interest from images that may not be easily apparent in the original (real space) spatial domain. For instance, in real-space, the AFM images contain noise emanating from the sample measurements which is visible along side the useful information. As a result, performing segmentation correctly in the spatial domain with thresholding techniques becomes a challenge (see examples in **Supporting Information Figure S2)**. In contrast, if we convert the image to another domain like the frequency domain, then it better isolates the frequencies pertaining to noise vs. those pertaining to the actual information we wish to capture.

To effectively utilize domain transforms for the task at hand, we define a workflow which is a sequence of operations executed on input AFM images to identify different domains in it. The workflow takes an AFM image as input and gives out two binary images each which signify the location of light and dark domains, as illustrated in **Figure 2**. The operations in a workflow include preprocessing, tiling, domain transform, post processing, and clustering described briefly next.

For preprocessing, we perform denoising and image normalization. Denoising techniques help reduce noise inherent in the AFM data and normalization is performed to scale the phase values which are obtained in degrees from the AFM experiment to quantize them into a fixed range comparable to natural images (0-255 intensity scale). See examples of images obtained pre-processing and post-denoising in **Supporting Information Section S.I**. Next, for the tiling stage, we define a region surrounding each pixel in the AFM image as a tile. A tile serves as a localized subset of the image, encompassing a neighborhood of pixels centered around each pixel. The advantages to analyzing tiles of each pixel in an AFM image are that they represent the spatial relationships and patterns within neighboring pixels which allows for the tile to be robust to noise and effective in representing the variability in individual pixel values. The size of the neighborhood or size of the tile is controlled by the *win factor*. The tile size is the win-factor multiplied by the AFM image size; tile size is a critical parameter as it directly influences the granularity of analysis and the level of detail captured within each tile. For more details about how to choose a tile size, we direct the reader to **Supporting Information Section S.II**. Due to the nature of tiles, it is not possible to formulate tiles for pixels in the boundary region as there may be no neighboring pixels in a tile. To prevent the loss of resolution some common techniques used in CNNs are padding. In padding, additional boundary pixels are added to the AFM image to prevent loss of resolution. However, doing so would introduce inconsistencies in texture. Therefore, we exclude boundary pixels from analysis in our workflow.

After generating tiles, domain transforms are applied to the pixel tiles within the AFM image. Each domain transform operates uniquely, resulting in varying output structures and each output structure has a unique meaning of what it represents. Therefore, the feature extraction from the transforms is specific to the transform used. The feature extraction is, in general, guided by how to compress and quantify the information from the domain transform output. Such compression is required because the transforms mostly output large two-dimensional (2D) matrices which are computationally intensive to work with. Statistical metrics like max, mean, variance, skew, and kurtosis of transform outputs, provide a compact representation of texture and are representable as low dimensional vectors which are computationally easier to work with. In short, the mean, variance, skew, and kurtosis of the transformed tile is used to represent the textural information of the pixel's tile. Collating each of the pixel's statistics would result in a three-dimensional (3D) grid of vectors, replacing the phase value with the calculated transform's statistics vector in the AFM image. We call such 3D grids as a "feature cube" (inspired by the 3D grid shape). After formulating the feature cube, we proceed with post-processing. Since the generated feature cube contains values spanning various ranges and scales of the statistical metrics, normalization becomes crucial to standardize the analysis, particularly for subsequent clustering processes. The workflow until creating the (normalized) final feature cube is illustrated in **Figure 3.**

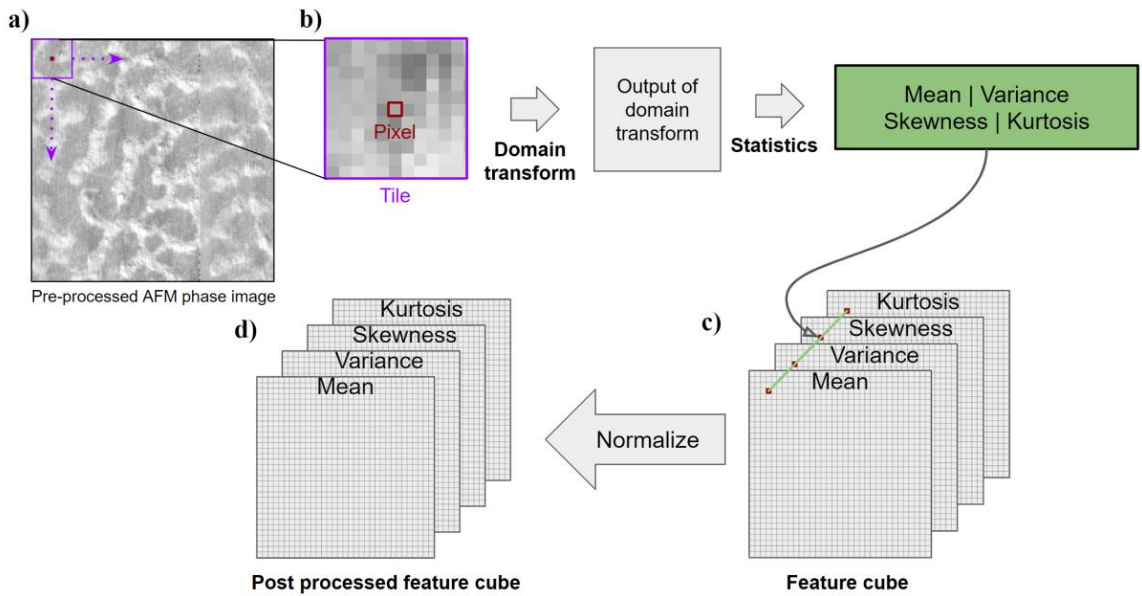

*Figure 3: Domain transform workflow. The (a) preprocessed AFM phase image is broken down into (b) tiles and domain transforms are applied on the (b) tiles. The domain transforms yield transformed representation of the tile which are easier for feature extraction. Statistical parameters like mean, variance,*

*skew, are kurtosis are used to reduce dimensionality and quantify textural features. These statistics are collated into a (c) feature cube which undergoes normalization in post-processing yielding a (d) post processed feature cube.*

Next, we will discuss the viability of the use of different domain transforms - discrete Fourier, discrete cosine, discrete wavelet, and Radon transforms - for our task of identifying phase separated polymer domains in AFM images taken from supramolecular block copolymer films.

**A.1. Discrete Fourier Transform (DFT)**

The two-dimensional discrete Fourier transform (2D DFT) is a mathematical operation that decomposes a two-dimensional image into its constituent sinusoidal waves; effectively the image goes from the spatial domain to the frequency domain where the image is represented by the frequency, amplitude, and phase information. [77] Mathematically, the 2D DFT of an image *f(m, n)* with dimensions M×N can be expressed as

$$F(k,l) = \frac{1}{MN} \sum_{m=0}^{M-1} \sum_{n=0}^{N-1} f(m,n) e^{-j2\pi\left(\frac{k}{M}m + \frac{l}{N}n\right)} \quad (1)$$

In Equation (1), *F(k,l)* denotes the transformed image in the frequency domain, while *k* and *l* represent the spatial frequencies in the vertical and horizontal directions, respectively. The spatial indices *m* and *n* correspond to the vertical and horizontal coordinates in the original image and *j* is the imaginary unit.

The 2D DFT yields a complex 2D matrix where the real part holds amplitude information, and the complex part holds the phase information. Recent work [64] has showed that amplitude information from the transform contains high potential to capture texture. This is because image texture is often characterized by repetitive patterns, which correspond to specific frequency components in the Fourier domain and these components are visible in the amplitude matrix. Therefore, applying statistics (mean, variance, skew, and kurtosis) helps us quantify the texture and provide a low dimensional representation which makes it easy for computing similarity or other desired metrics.

**A.2. Discrete Cosine Transform (DCT)**

The discrete cosine transform (DCT) is a mathematical operation used in signal processing with application to image and video compression. [78] DCT works similar to 2D DFT, where it converts the image from the spatial domain to frequency domain but unlike the 2D DFT, DCT decomposes images into constituent cosine waves rather than sinusoidal waves. Mathematically, the DCT of an image *f(m, n)* with dimensions M×N can be expressed as:

$$F(k,l) = \frac{2}{\sqrt{MN}} \sum_{m=0}^{M-1} \sum_{n=0}^{N-1} f(m,n) \cos\left[\frac{(2m+1)k\pi}{2M}\right] \cos\left[\frac{(2n+1)l\pi}{2N}\right] \quad (2)$$

In equation (2) *F(k,l)* denotes the transformed image in the frequency domain, while *k* and *l* represent the spatial frequencies in the vertical and horizontal directions, respectively. The spatial indices *m* and *n* correspond to the vertical and horizontal coordinates in the original image. As an imaginary unit is not involved in the transform the output of the transform is a real 2D matrix with amplitude and phase information. Recent work [79] has shown that the output of the transform has the potential to capture image texture information. This is because patterns in texture pose varying frequencies which are pronounced in DCT's output. Therefore, the mean, variance, skew and kurtosis are calculated on the 2D output matrix of DCT to capture texture information.

Both DCT and DFT work on the frequency domain and have similar characteristics of decomposition as explained above. Therefore, one could anticipate that the information on texture extracted from the two transforms could be similar; our results, as described later, confirm this is the case.

**A.3. Discrete Wavelet Transform (DWT)**

Discrete Wavelet Transforms (DWT) are mathematical techniques commonly used in image processing and image compression. [80] DWT decomposition works by convolving the image with low-pass and high-pass filters, which are specific to a chosen wavelet. Some commonly used wavelets include the Haar wavelet, symlets, and Daubechies wavelets, each with unique properties and decomposition characteristics.[81] For the task at hand, we use Haar wavelets and biorthogonal wavelets. The Haar wavelet is simple and efficient, ideal for capturing abrupt changes in an image. On the other hand, biorthogonal wavelets excel at detecting finer details and smoother trends. By experimenting with these two distinct types of wavelets, we aim to express the capabilities of DWT for texture analysis. DWT works by decomposing the image into coefficients of lower and higher frequencies. Lower frequencies are details in the image that have smooth variations and higher frequencies are details that have sharp or rapid variations. Furthermore, in the high frequencies, coefficients along vertical, horizontal, and diagonal directions of the image are captured. All this information is collated in one 2D matrix as illustrated in **Figure 4**.

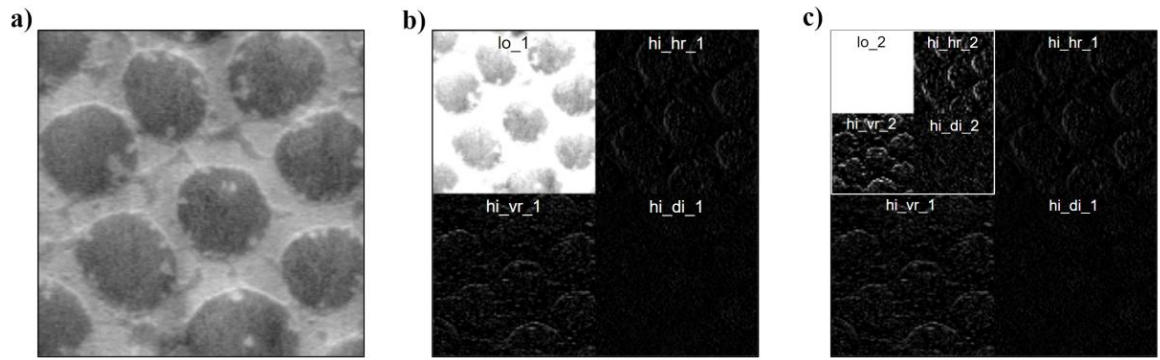

*Figure 4: Haar wavelet transform applied on the pre-processed AFM phase image shown in part (a). With one level of decomposition we obtain the (lo_1) low frequency coefficient, (hi_vr_1) vertical high frequency, (hi_hr_1) horizontal high frequency, and (hi_di_1) diagonal high frequency coefficient which are collated into one image in part (a). With decomposition level two, the (lo_1) previous low frequency coefficient from part (b) is decomposed further into 4 further sub images and all of them are collated in a similar fashion as part (b) yielding, (lo_2) low frequency coefficient, (hi_vr_2) vertical high frequency, (hi_hr_2) horizontal high frequency, and (hi_di_2) diagonal high frequency coefficient which are grouped into the*

*image shown in part (c). Note that the low frequency coefficients in (c) holds minimal detail and is approaching a stale state.*

By performing one level of decomposition on an image (Figure 4a) with DWT one yields 4 sub-images namely, low frequency coefficient, vertical high frequency, horizontal high frequency, and diagonal high frequency (Figure 4b). The size of these sub images is half of that of the input image because of which they could be collated as depicted in Figure 4b.

We note that multiple levels of decomposition could be performed where the low frequency image is further decomposed iteratively with DWT. (Figure 4c) The number of iterations in the decomposition process is referred to as the level of decomposition. The level of decomposition starts from 1 and goes to a value where the low frequency coefficients become stale. For the task at hand a level of decomposition more than 3 leads to stale low frequency coefficients. Unlike DFT and DCT, wavelets have complicated outputs with subsections in the matrix depicting various frequency coefficients. Therefore, it would be counterproductive to apply statistics to the entire output image. Instead, it is more advantageous to apply statistics to individual subsections of the output. By doing so, the mean, variance, skew and kurtosis quantify and capture the frequency patterns which depict properties of image texture with higher efficiency.

**A.4. Radon Transform**

Radon transform is a mathematical technique with extensive applications in medical imaging,[82] computer vision,[83] and hard material science [84]. It is commonly used in computed tomography (CT) scans for reconstructing images of human organs. CT scans work by using penetrating waves from different angles around the body to obtain projections. It has been shown that Radon transform can be applied to images to study image texture. [85] Radon transform works by integrating the intensity values of an image in a linear path which yields a vector called a 'projection'. Projections are taken at angles from 0 degrees to 180 degrees, which are then stacked to form a 2D matrix called the 'sinogram'. An illustration of projections and sinogram are presented in **Figure 5.**

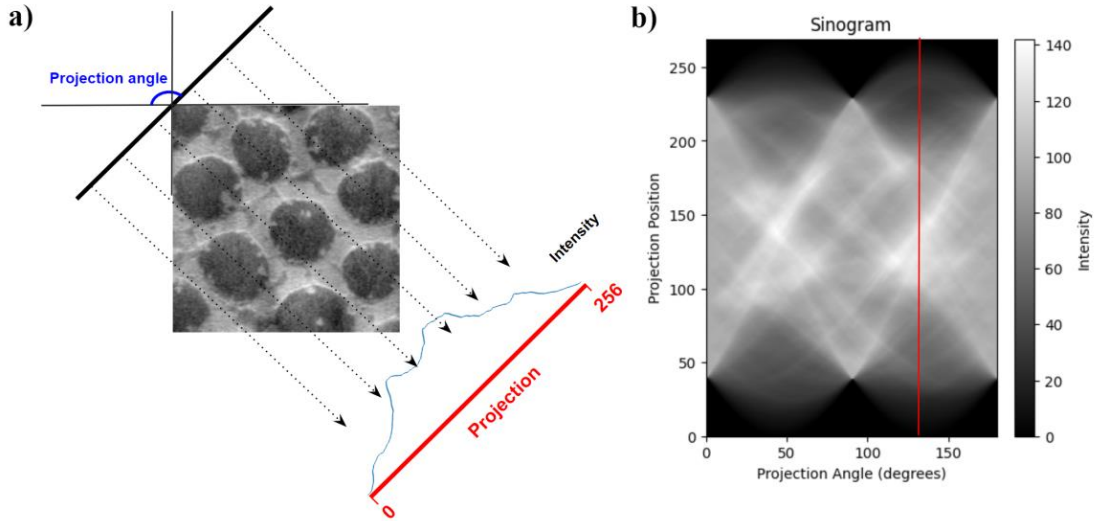

*Figure 5. Radon transform and visualization of sinogram.* The transform takes in the pre-processed AFM phase image (a) and calculates the integral of intensity values in the image along (dotted arrows) rays emanating from a projection angle with the AFM image. The projection angle varies from 0 deg to 180 deg and the integral of the rays is captured and stacked horizontally to form an image called the sinogram (b).

The sinogram, due to its integral information of intensities at various angles, tends to capture orientation and directional features from an image which could be quantified with the help of statistics mean, variance, skew, and kurtosis to extract textural features like directionality and regularity. Statistics also give a low dimensional representation of the sinogram making it easy for similarity calculations in clustering algorithms.

B.  **Deep Learning (DL) ResNet50**

*Table 2: Definitions of common terms used in the context of deep learning.*

| Term | Definition |
| --- | --- |
| Feature extraction | Process of transforming data into a set of more meaningful information or features. |
| Convolution | Mathematical operation used to either extract features or filter images. |
| Filter | Mathematical operation used on images to suppress or enhance desirable components. |
| Classification | Task of categorizing input data. |
| Index map | 2D matrix with indexes (0 or1) which signify presence of domain (light or dark). |
| Transfer learning | ML technique where knowledge gained from a training task is fine tunes to solve another task. |
| Unsupervised learning | Category of ML algorithms that extract pattern in data without the need of manual supervision. |
| Deep learning (DL) | A branch of ML where models contain multiple layers(deep) of neural networks. |
| Feature maps | Groups a 2D matrices obtained from the output of ResNet50 network stages. |
| ResNet50 | A pretrained DL model used for feature extraction. |

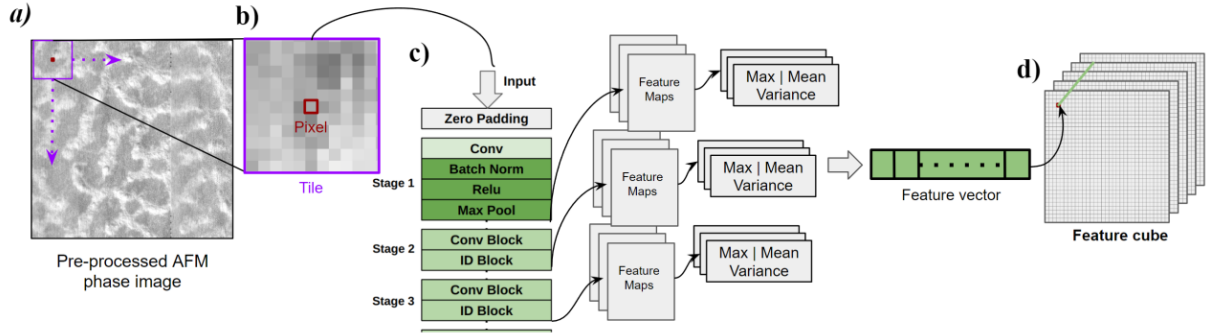

***Figure 6:*** *Schematic representation of ResNet50 workflow for domain segmentation till feature cube generation. The (a) pre-processed AFM phase image is divided into (b) tiles. The (b) tiles are fed into the (c) ResNet50 model which in turn generates feature maps. Statistical parameters - mean, max, and variance - are used to reduce the dimensions of the feature maps and these statistics are horizontally stacked in a feature vector. Repeating this process for all the tiles and stacking the feature vectors in spatial order, results in a (d) feature cube.*

Convolutional Neural Networks (CNN) is a deep learning method that is widely regarded as one of the most advanced approaches for image texture analysis and prediction. CNNs have multiple convolutional layers that are designed to learn and extract features from images. These layers contain multiple learnable filters which when convoluted over the input, detect features like patterns, edges, and corners, aggregating this information into a feature map. The learnable filters in CNN must be optimized to detect texture. This is done by training the model over millions of images with labels where it tries to correctly predict the image. CNN model training is supervised and a time and effort intensive process requiring large amounts of labeled data which often is not available in the soft materials characterization data. Therefore, if one wants to avoid training the CNN models from scratch themselves, they could use a pre-trained network which is trained on some task-specific image data or general images dataset and see how applicable that model is for their own task. One such pre-trained CNN is ResNet50 which was trained on a large dataset of images called ImageNet [86] consisting of 14 million images of various day to day objects; ResNet50 was trained to classify images with these objects. Recent work has shown that one could utilize various convolutional layers from models trained on ImageNet data to extract features from images. [66] Therefore, we could leverage ResNet50's ability to extract textural features from images (tiles in our workflow) and substitute domain transform with the ResNet50 model in the workflow. The resulting workflow is illustrated in **Figure 6**.

ResNet50 consists of multiple sets of convolutional layers where the level of understanding of features keeps reducing as we go deeper into the network. For the task at hand, we work with the output of the first 3 sets of convolutional layers to extract feature maps. The feature maps extracted have different sizes and represent textural information. Statistical parameters - mean, max, and variance - are calculated from the feature maps and are used to represent feature maps. Effectively this converts the feature map which is a 2D matrix to a vector of length 3. Grouping these vectors for each feature map gives us a feature vector which is collated to form the feature cube. Unlike in domain transform workflows, the feature cube is not normalized because CNNs inherently work on normalized data as a result the scale of values in the feature cube remains constant. It is important to note that despite the effectiveness of ResNet50, they still have limitations, one of which is the minimum input size which is restricted to 32x32 pixels; this restricts the granularity of analysis and the ability of the network to capture small scale features.

## C. Clustering

From the domain transforms and deep learning workflow, we typically obtain a three-dimensional feature cube. The first and second dimensions of the feature cube denote the spatial position on the raw image, while the third dimension contains the feature vector representing the textural information of the neighborhood surrounding that spatial coordinate (tile) indicated by the first two dimensions.

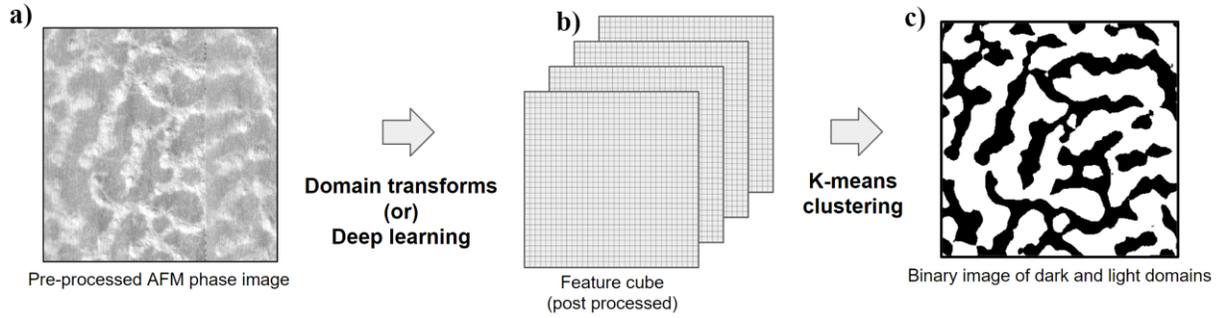

*Figure 7: **Schematics of clustering.** Domain transforms or deep learning workflows operate on (a) pre-processed AFM phase images to generate the (b) post-processed feature cube. Subsequently, the K-means clustering algorithm is applied on the post-processed feature cube to predict two distinct clusters. The resulting output from clustering is a (c) 2D image wherein each point is the index of a domain.*

An approach commonly used to classify regions from the feature vector involves training a supervised ML model. [45, 46, 66] However, supervised ML models need manual labeling of data for training the ML model. Furthermore, it makes it a data specific model, limiting the generalizability of the solution to other AFM images and workflow. To overcome this challenge, we chose an unsupervised approach to quantify the similarity of feature vectors within a feature cube and group them together using a process commonly known as clustering. Clustering, being an unsupervised ML technique, requires no labeled data or training. While numerous clustering techniques exist, each of their performance is data specific. A common and widely used clustering technique is k-means clustering. [87] K-means clustering works by partitioning all the feature vectors into clusters by iteratively assigning each feature vector to the nearest center of a cluster and updating the center of the clusters based on the mean of the feature vector assigned. Euclidean distance is calculated between feature vectors to assess the distance between feature vectors in this process. Upon completion of the clustering process, ideally the feature vectors in each cluster would have similar textures. The number of clusters in k-means is a hyper-parameter, and for the task at hand, it was observed that two clusters best represent the light and dark domains for the pairs of described domain transforms and statistics from **section II.A** and **II.B**. One may also note that based on the nature of the problem and transforms multiple clusters may be observed to represent a single texture in this case the multiple indexes must groups as one. By replacing the feature vectors in the feature cube by indexes which signify light or dark domains, we have a binary 2D matrix (index map) signifying the position of the domains on the raw image. (**Figure 7**) This 2D matrix (index map) is then used to calculate the domain size distribution (**Figure 8**).

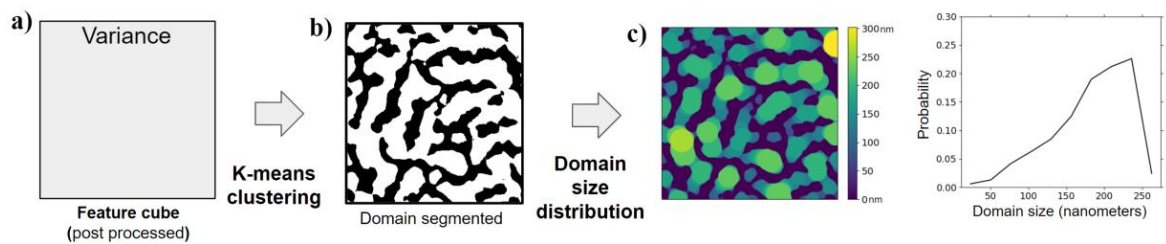

*Figure 8: K-means clustering is applied on the (a) feature cube resulting from DFT (best transform for the specific goal of this paper) and variance statistics (as shown in **Figure 3**). This results in (b) an index map which represents the spatial position of domains within the image. We then use preexisting algorithms for domain size distribution [88] on the (b) index map for light domains to generate (c) a heat map that depicts the radius of the largest circle that could overlap that pixel. Also, in the heat map the pixels of the dark domain are zero when calculating on light domain. We repeat the process for the index map of the dark domain too. Using the scale bar in the metadata associated with the AFM image, from the heat map we calculate the domain size distribution in the appropriate distance units (nm for the AFM images we use).*

D. Metrics Used for Evaluating Model's Performance

Performance metrics are key to quantitatively measure the accuracy and efficacy of any workflow. The most widely used performance metrics for image segmentation problems include accuracy, Dice coefficient, and intersection of union (IoU).

Accuracy quantifies the correctness of the index map and is defined as the fraction of number of correct index predictions divided by the total number of indexes in the index map.

Dice similarity coefficient [89] is a useful metric used to measure the spatial overlap of domains from the predicted and ground truth images, Dice score for one domain is calculated as twice the number of pixels with a common index (of the observing domain) in both the predicted and manually segmented images, divided by the total number of predicted and manually segmented pixels containing the index of the observing domain.

The overall Dice score for the complete workflow is defined as the weighted average of Dice score for each domain calculated as,

$$Dicescore = \frac{1}{2}\sum_{i=0}^{1} Dice_i \quad (3)$$

$$Dice_i = \frac{2 \times |A_i \cap B_i|}{|A_i| + |B_i|} \quad i = 0,1 (light, dark)$$

$$A_i = set of pixels \in predicted index map with index i$$

$$B_i = set of pixels \in maunally segmented index map with index i$$

$$|.| = cardinality of set$$

The Jaccard index or intersection of union (IoU) [83], like Dice, measures the spatial overlap of domains from the predicted and ground truth images. IoU for the entire workflow is calculated as the weighted average of IoU calculated for each domain where IoU of a domain is calculated as shown below:

$$IoU = \frac{1}{2}\sum_{i=0}^{1} IoU_i \quad (4)$$

$$IoU_i = \frac{|A_i \cap B_i|}{|A_i \cup B_i|} \quad i = 0,1 (light, dark)$$

$$A_i = set of pixels \in predicted index map with index i$$

$$B_i = set of pixels \in maunally segmented index map with index i$$

$$|.| = cardinality of set$$

**III. Experimentally Obtained AFM Image Datasets**

Supramolecular block copolymer poly(oligo(ethylene glycol) methyl ether methacrylate)-sb-polystyrene (POEGMA-sb-PS) was fabricated by blending two types of homopolymers: first homopolymer is diaminotriazine (DAT) functional POEGMA (POEGMA-DAT) and the second homopolymer is thymine (Thy) functional PS (PS-Thy). As the focus of this paper is on the machine learning workflow development, we do not present any details of the synthesis of these homopolymers which Gu and coworkers will present in a future publication focused on the synthesis of these molecules. Homopolymers POEGMA-DAT and PS-Thy in 1:1 molar ratio were dissolved in anhydrous toluene at room temperature to achieve a solution with polymer concentration of 20 mg/ml; the two homopolymers can associate to form supramolecular block copolymer POEGMA-sb-PS. By using varying molecular weights of POEGMA-DAT and PS-Thy homopolymers we achieved supramolecular block copolymers POEGMA-sb-PS with varying total chain lengths and block ratios.

Finally, thin films of POEGMA-sb-PS were spin-cast onto silicon wafers at 2000 revolutions per minute (rpm) for one minute for characterization. AFM images of the films of POEGMA-sb-PS were acquired on an Asylum Jupiter AFM microscope in AC-air mode. The dataset we use for the development of the machine learning workflow has 144 AFM images in total, imaged from 16 samples of POEGMA-sb-PS obtained by using 7K and 10K Da PS blocks and 5K, 8K, 12K, 14K, 16K, 20K, 23K and 26K Da POEGMA blocks. Within each sample, the domains have relatively similar shapes. Across samples the domains vary drastically in size and shape.

Each of the 144 AFM images contains distinct texture/domain patches of varying sizes and shapes, as depicted with representative images shown in Figure 1. These images in their raw form present phase angles which are quantized to intensity values for pixels and have dimensions of 384 pixels x 384 pixels in each image. Additionally, metadata is extracted from AFM images. The metadata contains details on measurement settings and length scales where the latter is helpful in mapping a pixel in an AFM image to length in real space (in nanometers). Visually the different texture patches in AFM images could be perceived as relatively light and dark regions, representing the two different domains. The AFM images in our dataset inherit noise stemming from environmental factors and experimental conditions. The prevalent types of noise observed include line, random, and scar noise. [90] Utilizing prebuilt denoising algorithms tailored for AFM images enables effective noise reduction, yet some residual defects remain post-denoising (see examples in **Supporting Information Section S.I.**) We note that our workflow can accommodate and tolerate these remaining imperfections.

**IV. Results**

We first assess the domain segmentation workflow with the various transforms and ResNet to evaluate their performance in terms of accuracy, consistency, and robustness handling our AFM image data. We do not assess the effectiveness of the domain size quantification methods, as we are using existing approaches (porespy python package) for this step [88]. In principle the user could choose to develop their own in-house codes for any other desired domain quantification (e.g., percolation, tortuosity) after the domain segmentation task on the AFM image is complete.

**A.     Assessment of Various Transforms and ResNet for Domain Segmentation**

In **Figure 9** we present representative results of domain segmentation conducted with all transforms (described in **section II.A**) and ResNet50 (described in **section II.B**). Visually, the results of domain segmentation from discrete Fourier transform (DFT), discrete cosine transform (DCT), and Radon transform show the best match to the original AFM image. As described in **section II.A**, the similarity between DFT and DCT leads to their workflows having similar domain segmentations outputs. Additionally, we see similarity in the output of Radon transform workflow and the results from DFT and DCT. However, there are minor differences in prediction between Radon transform and DFT which are illustrated in **Supporting Information Figure S7.**

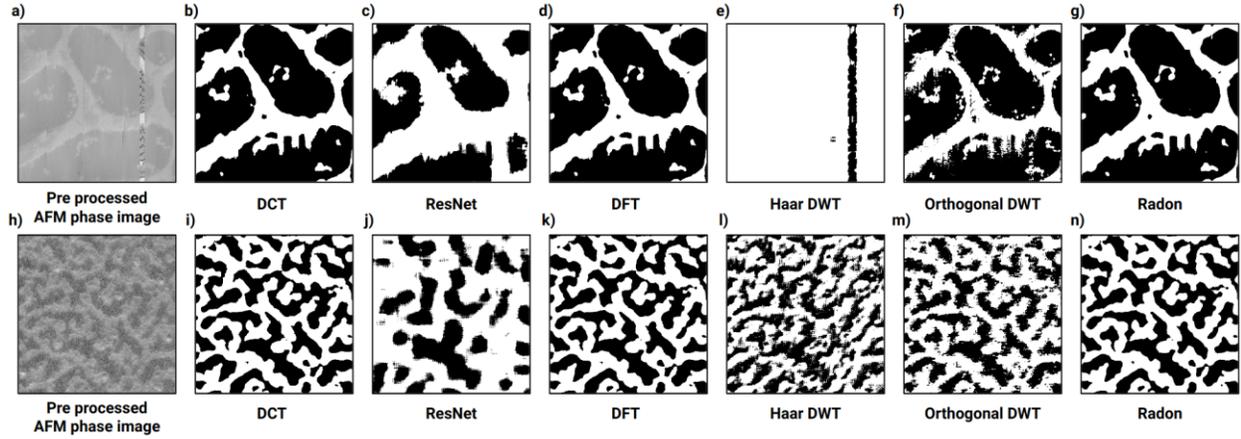

*Figure 9: For two representative AFM images (a and h), the results of domain segmentation from various transforms and ResNet50 (deep learning model) are shown. The type of approach is denoted in the text below the images. Visually we can see that DFT, DCT, and Radon workflows perform the best by producing a binary image that is most similar in pattern to the original pre-processed AFM phase image.*

In the next few sub-sections, we will describe results from DFT workflow in more detail and we direct the reader to take these discussions on DFT workflow to be representative of DCT workflows as well. For results from DWT (Haar and bi-orthogonal wavelets), we direct the reader to the **Supporting Information Section S. III.**

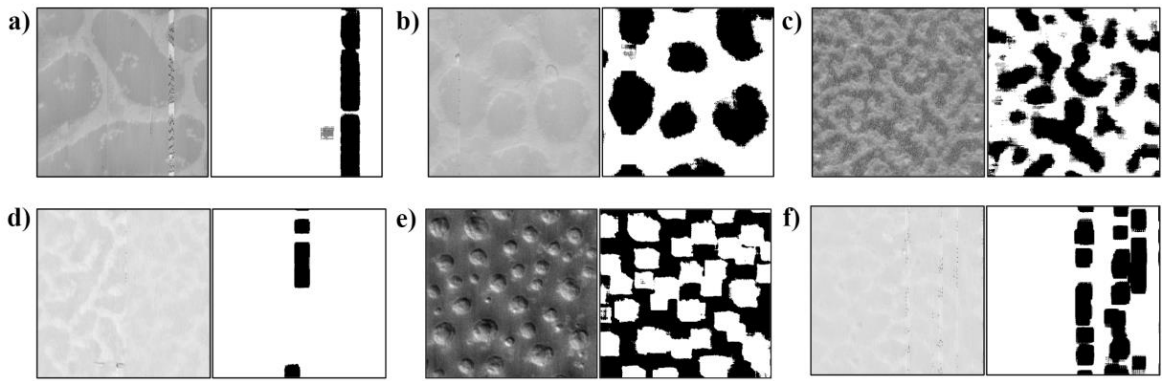

*Figure 10: (a-f, right) Pairs of domain segmented images, corresponding to (a-f, left) preprocessed AFM phase images when processed through **ResNet50** workflow. The results exhibit noise (a, d, f) and accuracy of predictions in domain segmentation is moderate (d, c, e).*

Before we discuss the DFT results in more detail, we briefly describe the performance of ResNet50 because such deep learning workflows are widely regarded as state of the art in image segmentation. From Figure 9c and 9j, we observe that the domain segmentation using ResNet50-based workflow does not perform as well as the DFT workflow for our AFM images. We provide additional examples of poor domain segmentation results from ResNet50 model in **Figure 10**. The overall poor performance of ResNet50 for the domain segmentation task as hand could be attributed to two factors: the choice tile size and the need for transfer learning.

The reader may recall that tile size is controlled by the win factor which is an important parameter that regulates the transforms or deep learning models like ResNet50 to look for textural features of a particular spatial size. We found a win factor of 0.03 to work best for our dataset (see **Supporting Information Section S.II**). This win factor yields a tile size of 12 pixels x 12 pixels whereas ResNet50 has a fixed minimum of 32 pixels x 32 pixels. This causes the ResNet workflows segmentation resolution to be low (Figure 10b, 10c, and 10e) and predictions prone to noise (Figure 10a, 10d, and 10f). Second, the poor domain segmentation with higher sensitivity to noise as compared to DFT results suggest we need to improve the ResNet50 workflow. In **Supporting Information Section S.V.**, we present additional steps that we took as we experimented with improving the ResNet50 performance. As we did not see significant improvement with ResNet50 or other domain transform, we declared DFT and DCT workflows the best of all the approaches we tested here.

**Figure 11** shows a few more examples that demonstrate visually that DFT workflow captures the domain segmentation for six more AFM images correctly even in the presence of noise (see the presence of a line in Figure 11e, dots in Figure 11f and blur in Figure 11b and 11c). DFT workflow also offers high computational speed and low memory usage, with processing speeds of 7 images per minute benchmarked on Intel i9-12900H CPU and 16GB of DDR5 RAM.

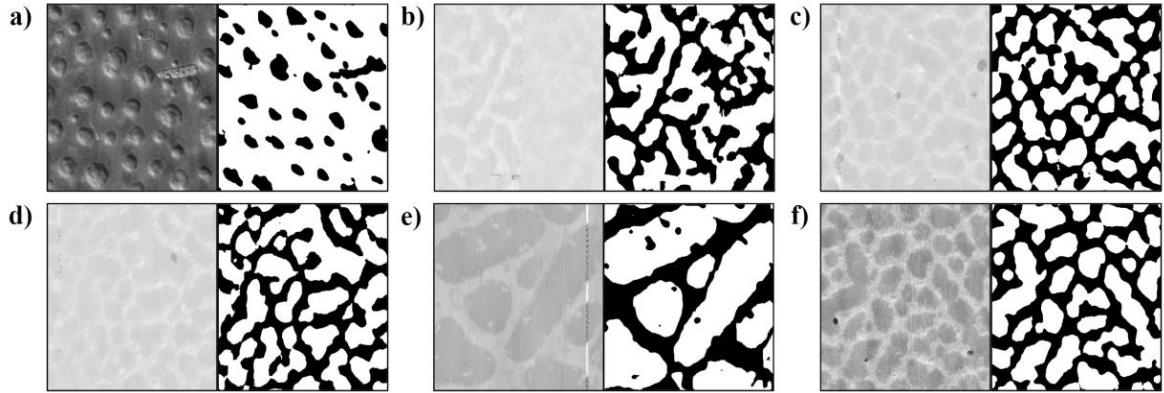

*Figure 11: Segmentation results using DFT workflow:* In each of the six parts of this figure, we show the (left) preprocessed AFM phase images and (right) the completed domain segmentation from the DFT workflow with variance statistics. If the spatial organization of the dark or white regions in the right images correspond to the patterns in the AFM image, then that would be a successful prediction.

**B. Analyzing the Statistical Features in DFT Workflow for Best Feature Extraction.**

For the DFT workflow, of all the statistical features – mean, variance, skewness, kurtosis – we find that the *variance* is the best metric for dimensionality reduction as it distinctly represents the two domains from the feature map. **Figure 12** presents comparisons for the domain segmentation results when using the various statistical features for dimensionality reduction. For input AFM images in Figure 12a and 12d, we see domain segmented images when DFT was applied using all statistical features - mean, variance, skew, and kurtosis- in Figure 12b and 12e, respectively. In Figure 12b we see the algorithm identifies the domain boundaries but not the domains themselves whereas Figure 12e tries to predict the bulk of the domain. In contrast, when DFT was applied only using the variance, (Figures 12c and 12f) the domain segmented images correctly capture the domains in the original AFM images.

Upon analyzing the correlation between the statistics (Figure 12g), we find that the kurtosis and skewness are perfectly correlated, there is high correlation between mean, skewness, and kurtosis, and a lack of correlation of variance with any of the other three statistical features. These correlations among other features or lack thereof in case of variance warranted us to declare variance as the best metric to uniquely identify both the domains.

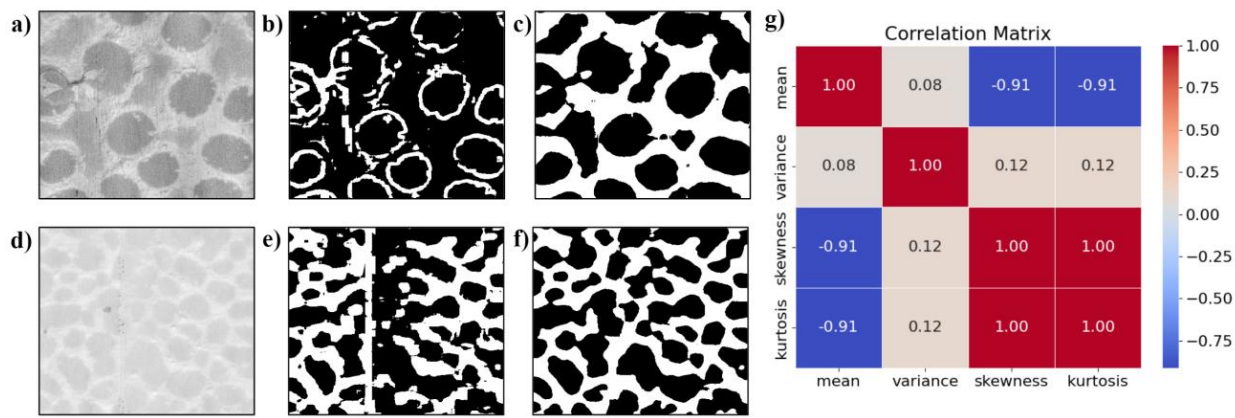

*Figure 12: Analysis of statistics used in DFT workflow. (a, d) are two preprocessed AFM phase images used for DFT workflow. Parts (b) and (e) are the domain segmented images when DFT was applied using all statistical features - mean, variance, skew, and kurtosis - for images in parts (a) and (d), respectively. part c and f are the domain segmented images when DFT was applied using only one statistical feature variance for images in parts a and d, respectively. Part (g) presents the correlation matrices of the 4 statistics that show the mean, skewness, and kurtosis are highly correlated amongst themselves, and variance is mostly independent from the rest.*

## C. Quantification of Segmentation Performance of DFT Workflow with Variance Statistics

So far, our assessment of the DFT workflow with variance statistics has been purely qualitative using our visual sense. To quantify the performance of DFT workflow that visually looked superior to other methods, we created a test dataset selecting raw AFM images from the dataset with unique texture patterns arising from different polymer systems. To analyze the performance of domain segmentation, we manually segmented the test dataset images with an image annotation tool [91] to generate a (manually) annotated index map. Using various performance metrics (as defined in **section II.D**) we compare the manually annotated index map to the predicted index map from DFT workflow using variance as feature (**Figure 13**).

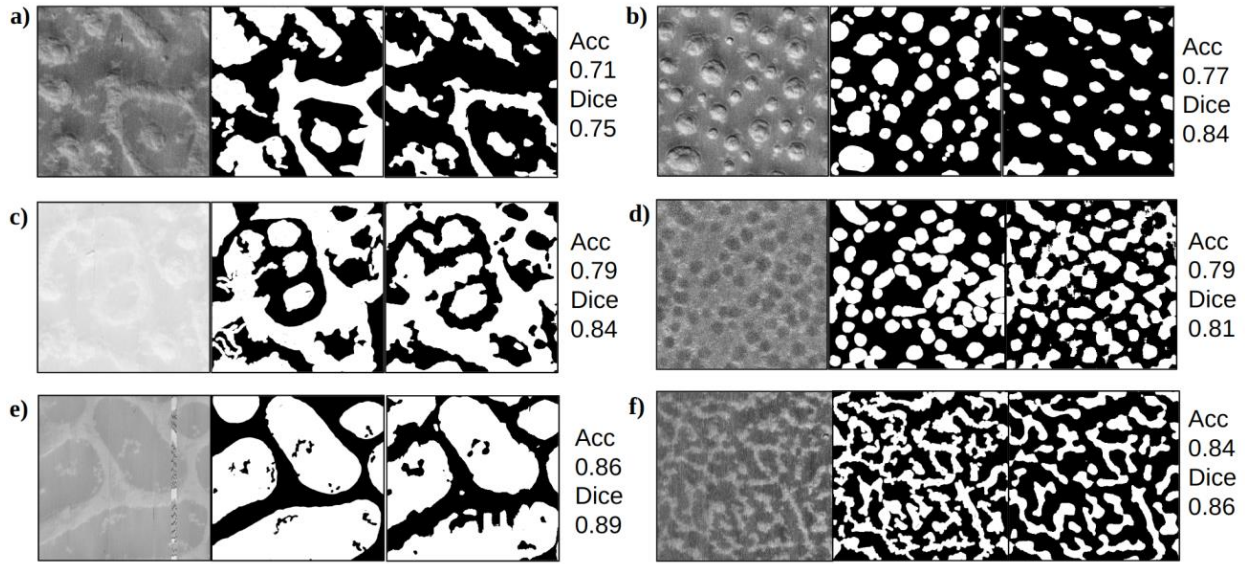

*Figure 13:* *Results of AFM images segmentation using DFT workflow with variance statistics. For six preprocessed AFM phase images (left) from testing dataset we provide the Acc (accuracy) and Dice score for the comparison of domain segmentations using DFT workflow with variance statistics (right panel) versus 'ground truth' manual segmentation (middle panel).*

Before we discuss the results of this quantified performance, it is critical for us to highlight the limitations of evaluating the predicted data in this manner. First, manual annotations have a precision vs. time tradeoff where higher precision in annotations consume more time to annotate even for a small test dataset. Balancing precision and time leads to some compromise in precision which makes the process prone to errors in the boundary regions of domains. Second, manual annotations are subjective (i.e., person's perspective) leading to inconsistencies especially when two domains look similar to one person but different to another person. Therefore, the similarity metric calculated from the manual annotations is a good representation to measure the bulk of the workflow's performance but does not give us an accurate quantitative metric for each prediction as the manual annotations themselves could be flawed due to manual segmentation.

For all the metrics described in **section II.D**, namely, accuracy, Dice coefficient and IoU, the value ranges from 0 to 1 where 1 is the most accurate and 0 is the least. In **Table 3**, we present the metrics of similarity between the predicted domain segmentation with the manual segmentation of AFM images in the test dataset for the 'best' workflow (i.e. DFT workflow + variance statistics + k-means clustering). From the results for specific images in Figure 13 and the collective performance metrics in Table 3 we are able to show that the presented workflow using DFT with variance statistics and k-means clustering performs quite well.

**Table 3:** Average performance metrics of segmentation on all images of test dataset with DFT workflow and variance statistics.

| Metric | Value |
|---|---|
| Accuracy | 0.81 |
| Jaccard Similarity | 0.76 |
| Dice Coefficient | 0.85 |

### D.    Calculated Domain Size Information from the Segmented AFM Images

In **Supporting Information Figure S12,** we present the distributions of light and dark domain sizes for 15 selected AFM images in our dataset of 144 images. We observe that the quantitative predictions of domain sizes are consistent with human interpretation of the domain sizes (comparing with the help of the scale bar present in the image). In **Supporting Information Figure S13**, we present cumulative results of all AFM images from the various polymer films along with the average, standard deviation, maximum, and minimum of light and dark domain sizes observed in the films. These statistics are calculated on the average of domain sizes observed in AFM images captured across a polymer film sample. Thus, these statistics collate the information of domain sizes observed in a film which helps us draw various conclusions on the workflow's performance and how the composition affects domain sizes in the polymer films. For instance, the small standard deviation of domain sizes from multiple AFM images captured from different regions of the same polymer film show that the domain size distribution across the sample is similar; this also indicates the workflow's consistency in predictions. Furthermore, the light and dark domain sizes do not have a linear nor an inverse relationship (in other words, we cannot guess the light domains' sizes from the dark domains' sizes) therefore it is essential to analyze both the light and dark domain sizes, independently. We also found a non-monotonic trend in the light and dark domain sizes for *PS-Thy 7K and PS-Thy 10K* with increase in molecular weight of POEGMA blocks. In a future publication, Gu and coworkers will describe the fundamental insights into these trends in detail.

## V. Conclusion

Analysis of features of AFM images with polymer blends has traditionally been a time intensive manual task which inherits inconsistencies, biases, and errors. Our ML based workflow automates identification and quantification of domain sizes in AFM images of polymer blends. Our workflow comprised of DFT or DCT domain transforms with variance statistics and k-means clustering worked best in segmenting AFM images containing two types of domains from phase separating polymers. Besides visual evaluation, we quantified the performance of our workflow by calculating overlap in our workflow prediction against the manually annotated images for a small test dataset. We found high accuracy in overlap captured by Dice coefficient of ~0.85 on average on all of the test dataset. Further, the predictions accuracy is consistent across domains of different shapes and textures which proves the generalizability of the solution.

This paper also shows that efficient problem formulation enables the nature of the workflow to remain unsupervised, which significantly reduces the traditional labor-intensive and subjective nature of manual interpretation of AFM images in soft materials science. As a result, it speeds up the development process and brings uniformity to analysis.

Lastly, our paper is meant to serve as a guide for readers in polymer science and soft materials who may wish to extend the discussed methods to other microscopy images captured from soft materials. To transfer the knowledge in a precise manner we present a short summary of the various transforms and ML approaches we tested along with our experiences as we tested them for the task we had at hand.

**Table 4:** Summary of transforms and machine learning approaches and our experiences during testing along with parameters that the user has to work with.

| Approach | Our Experience | Parameters That User Selects |
|---|---|---|
| Haar wavelet | • Light on computational intensity<br>• Works on gray scale images.<br>• Can adjust the level of decomposition.<br>• Requires no training.<br>• Emphasizes gradients.<br>• Explains and quantifies features.<br>• Works well on small tile sizes. | • Level of decomposition<br>• Window size<br>• Number of clusters or classes |
| Biorthogonal wavelet | • Light on computational intensity<br>• Works on gray scale images.<br>• Can adjust the level of decomposition.<br>• Requires no training.<br>• Works well on small tile sizes. | • Type of wavelet<br>• Window size<br>• Start level of decomposition<br>• Level of decomposition<br>• Number of clusters or classes |
| Fourier transform | • Light on computational intensity<br>• Works on gray scale images.<br>• Requires no training.<br>• Explains high vs low frequencies.<br>• Works well on small tile sizes. | • Window size<br>• Number of clusters or classes |
| Radon transform | • Light on computational intensity<br>• Works on gray scale images.<br>• Easy to extract directional features & orientations.<br>• Requires no training.<br>• Works well on small tile sizes. | • Type of wavelet<br>• Window size<br>• Starting level of decomposition<br>• Level of decomposition<br>• Number of clusters or classes |
| Deep learning ResNet50 | • Able to generate large dimensional features.<br>• Able to classify complex details.<br>• Works on RGB images.<br>• Requires no training if use as is makes sense.<br>• Compute relatively slower.<br>• Minimum tile size of 32 pixels x 32pixels.<br>• Features are not explainable due to black box implementation.<br>• Highly susceptible to noise.<br>• Inputs must be normalized. | • Window size min 32 pixels x 32 pixels.<br>• Layer selection for feature extraction.<br>• Selection of weights for model.<br>• Number of clusters or classes. |

## Open-source Code availability

We have made the Python program files open-source and public on this website.

https://github.com/arthijayaraman-lab/Automated-Atomic-Force-Microscopy-Image-Analysis

## Acknowledgements

A.J. and A.P. are grateful for financial support from Multi University Research Initiative (MURI) from the Army Research Office, Award Number W911NF2310260. Y. W. and X. G. are grateful for financial support from the Department of Energy under the award number of DE-SC0024432. A portion of this work was done at the Molecular Foundry, which is supported by the Office of Science, Office of Basic Energy Sciences, of the U.S. Department of Energy under Contract No. DE-AC02-05CH11231.

**SUPPORTING INFORMATION**

# Machine Learning for Analyzing Atomic Force Microscopy (AFM) Images Generated from Polymer Blends

Aanish Paruchuri [1], Yunfei Wang[2], Xiaodan Gu[2], Arthi Jayaraman [3,4,5] *

1. Master of Science in Data Science Program, University of Delaware, Newark DE 19713
2. School of Polymer Science and Engineering, 118 College Drive, #5050 University of Southern Mississippi, Hattiesburg, MS 39406
3. Department of Chemical and Biomolecular Engineering, 150 Academy St, University of Delaware, Newark DE 19713
4. Department of Materials Science and Engineering, University of Delaware, Newark DE 19713
5. Data Science Institute, University of Delaware, Newark DE, 19713

* Corresponding authors **arthij@udel.edu**

# S.I. Dataset

### A. Link to raw data

Raw data used in this study has been deposited in Zenodo DOI:
https://zenodo.org/doi/10.5281/zenodo.11179874

### B. Visualization of noise in dataset prior and post denoising

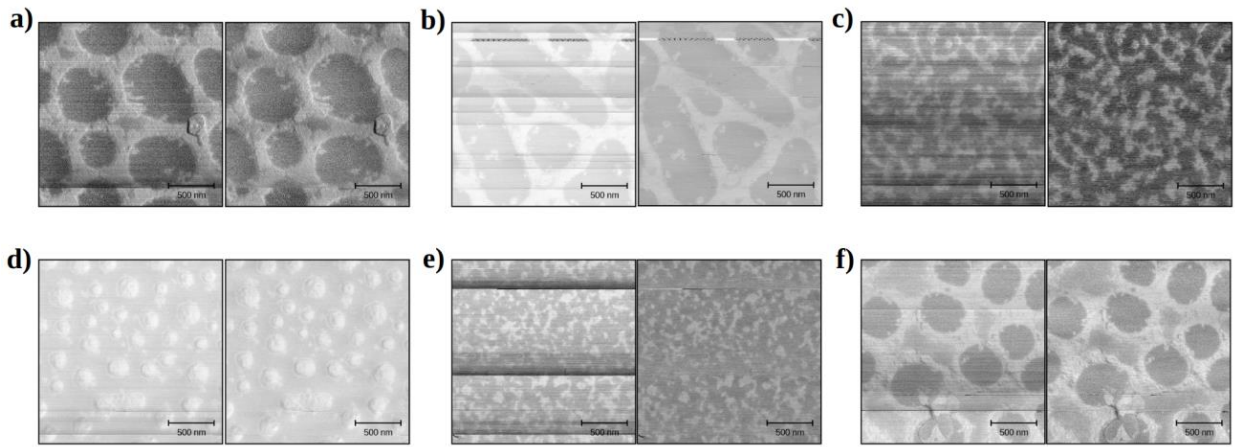

*Figure S1:* Image pre-processing results. (a-f) For the presented six AFM images, the image on the left corresponds to the noisy AFM phase measurement and the image on the right corresponds to the denoised image. Image denoising operations are performed by Gwyddion software [1].

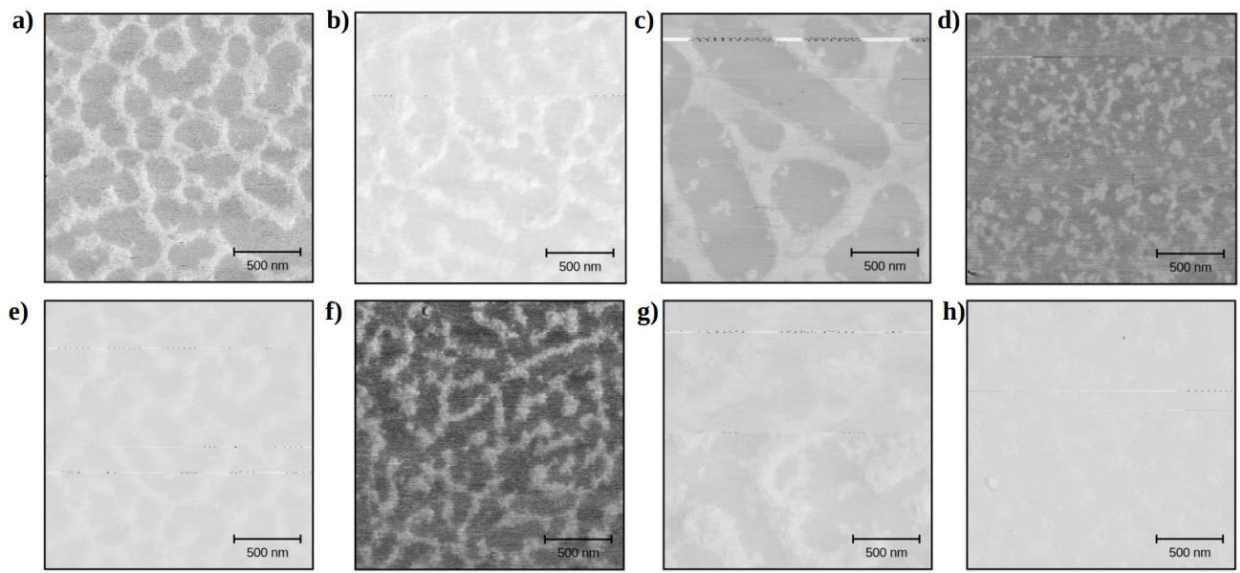

*Figure S2:* (a-h) Pre-processed AFM phase images after denoising with Gwyddion[1] software; these images are shown to demonstrate that even after denoising these images retain some noise (visible defects).

## C. Thresholding techniques applied on dataset

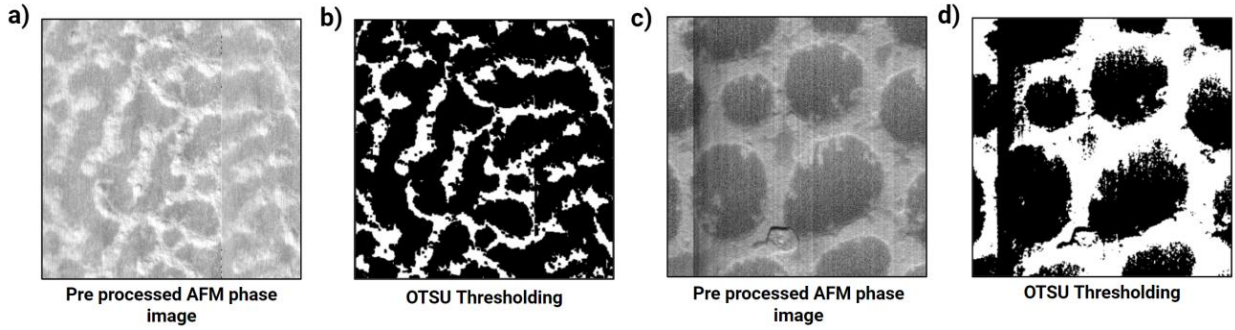

*Figure S3:* OTSU thresholding algorithm applied on (a,c) pre-processed AFM phase images after denoising. (b,d) are their respective outputs when OTSU is applied locally on patches of the pre-processed AFM phase images and the denoised with image morphology operation. Thresholding fails in (d) due to the line noise and (b) lacks consistency in predicting continuity.

# S.II. Tiles and Win Factor

## A. How to choose the best *win factor*

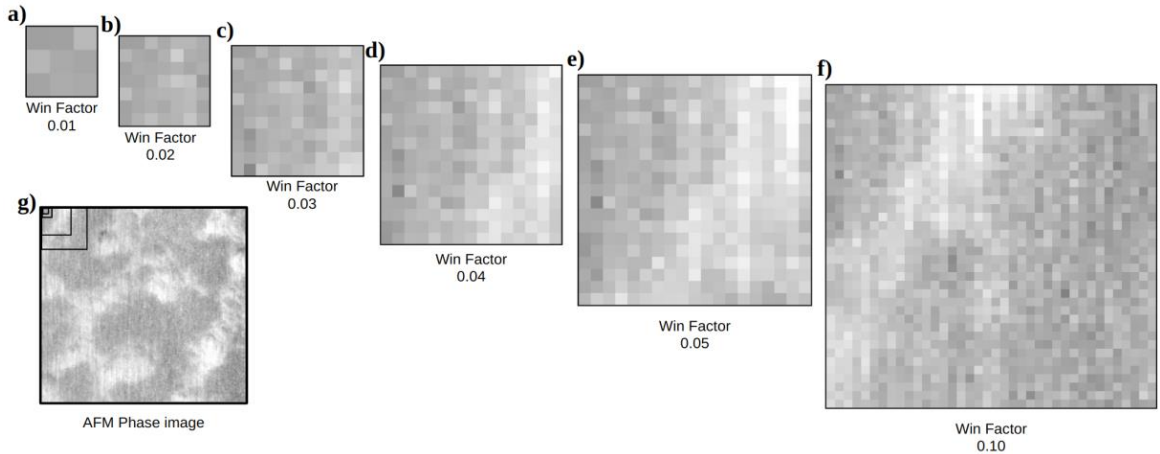

*Figure S4:* (a-f) Samples of tiles with varying win factor from 0.01 to 0.10 generated from one pre-processed AFM phase image shown in part (g). To find the best tile size, we increase the win factor iteratively and choose the minimum size that can distinguish between light and dark domains in the AFM image. We can see that in parts (a) and (b) with low win factors 0.01 and 0.02 one cannot distinguish between the two domains. We notice that in part (c) the win factor of 0.03 presents a visual difference in light and dark domains. Therefore, we choose this as the best win factor. To confirm the details are of domains and not noise, we also generate for the reader the images in parts (d-f) with larger win factors.

## B. Pseudo code of generating *tiles* from input image with given *win factor*

A raw image is sub-sampled into tiles with the following workflow using steps below:

$$tile\_width = input\_img\_width * win\_factor \quad (1)$$

$$tile\_height = input\_img\_height * win\_factor \tag{2}$$

$$T\_ij = input\_img\_img[i - \frac{tile\_width}{2} : i + \frac{tile\_width}{2}, j - \frac{tile\_height}{2} : j + \frac{tile\_height}{2}]$$

*where,*

$$i = (\frac{tile\_width}{2}), (\frac{tile\_width}{2} + stride), (\frac{tile\_width}{2} + 2*stride), \ldots, (input\_img\_height - \frac{tile\_width}{2})$$

$$j = (\frac{tile\_height}{2}), (\frac{tile\_height}{2} + stride), (\frac{tile\_height}{2} + 2*stride), \ldots, (input\_img\_height - \frac{tile\_height}{2}) \tag{3}$$

Variables *stride* and *win factor* have a direct impact on feature extraction and are also responsible for the resolution of domain segmentation output. *Stride* has values starting at 1. As we increase *stride*, it decreases the resolution in prediction of domain segmentation output. The *win factor* is responsible for the tile size and a large tile size results in more boundary pixels excluded from analysis (the latter is described in main manuscript's **section II A).**

## S.III. Discrete Wavelet Transform (DWT)

### A. Performance and characteristics of different wavelet types in DWT workflow

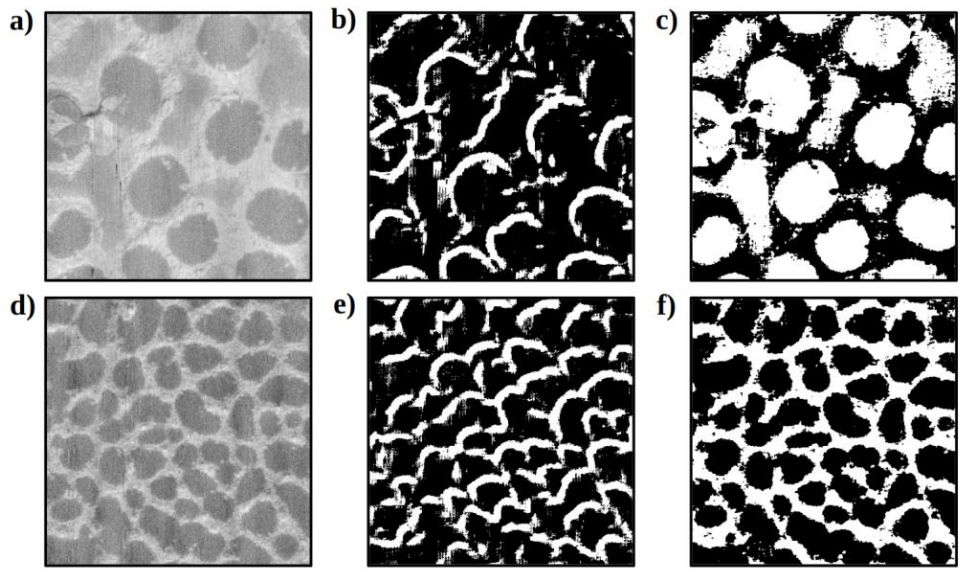

*Figure S5:* Parts *(a) and (d) are two preprocessed AFM phase images and parts (b, c) and parts (e, f) are their domain segmentation outputs, respectively. The DWT workflow's domain segmentation outputs from (b, e) Haar wavelet and (c, f) biorthogonal wavelets are shown. Depending on the convolutional filters inherited by the type of wavelet, each has a different characteristic of decomposition. (b, e) Haar wavelets focus more on capturing larger gradients whereas (c, f) biorthogonal wavelets capture more continuous gradients. As a result, we get to see different domain segmentations in both the cases (b, e) and (c, f).*

### B. Extension of DWT workflow on other literature datasets: scope and opportunities

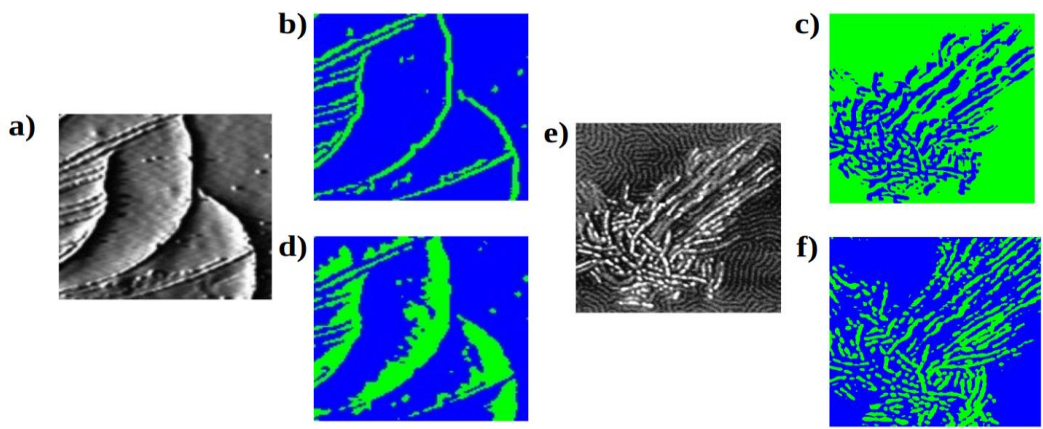

*Figure S6: DWT methods applied to other AFM images (parts a and e) adapted with permission from [2, 3] Copyright 2001 American Chemical Society. The images in (b, c) and (d, f) are the corresponding Haar*

and bioorthogonal domain segmentations for parts a and e, respectively. The domain segmentations are useful to study fibril like patterns in AFM images where features like length, directionality, and orientation of fibrils are of interest.

## S.IV. Radon Transform

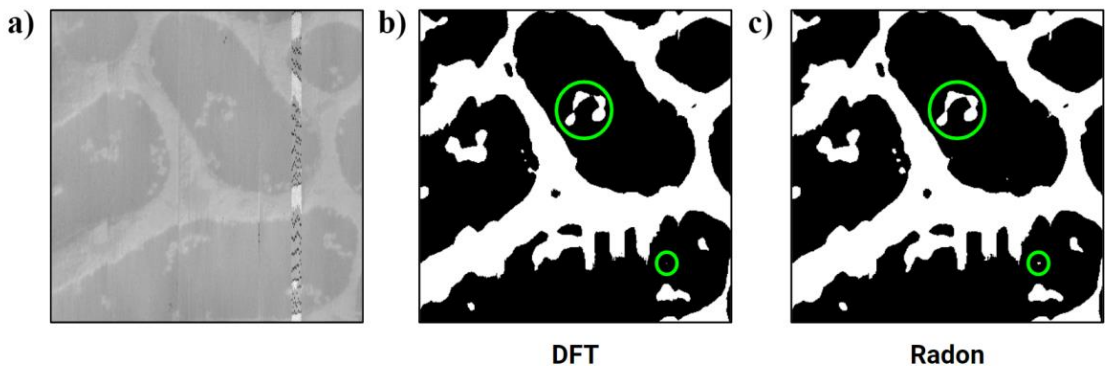

***Figure S7:*** *DFT and Radon workflow applied on (a) a pre-processed AFM phase image resulting in (b, c) domain segmentations. In parts (b) and (c) the green circles highlight the minor differences in segmentation obtained from these two workflows.*

## S.V. ResNet50: Methods experimented to improve ResNet50 performance

### 1. Methods to improve performance on noisy data.

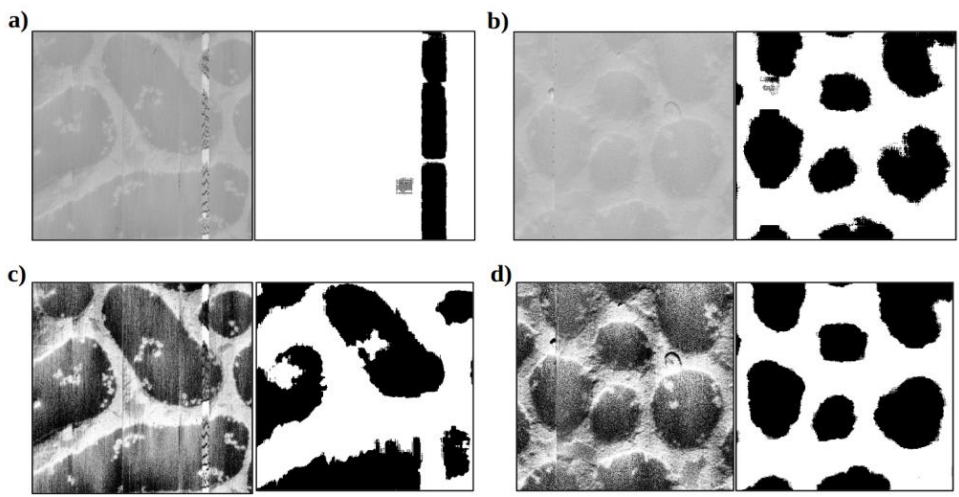

***Figure S8:*** *ResNet50 workflow applied on pre-processed AFM phase images shown on the left in parts (a) and (b) without histogram equalization results in the right image in parts (a) and (b). As we can see the results are prone to noise in the pre-processed AFM image as ResNet50 is sensitive to scale and outliers in the image. Applying histogram equalization to the preprocessed AFM phase images before using the ResNet50 workflow shows tremendous improvement in the results as shown in parts (c) and (d).*

## 2. Methods to address minimum tile size.

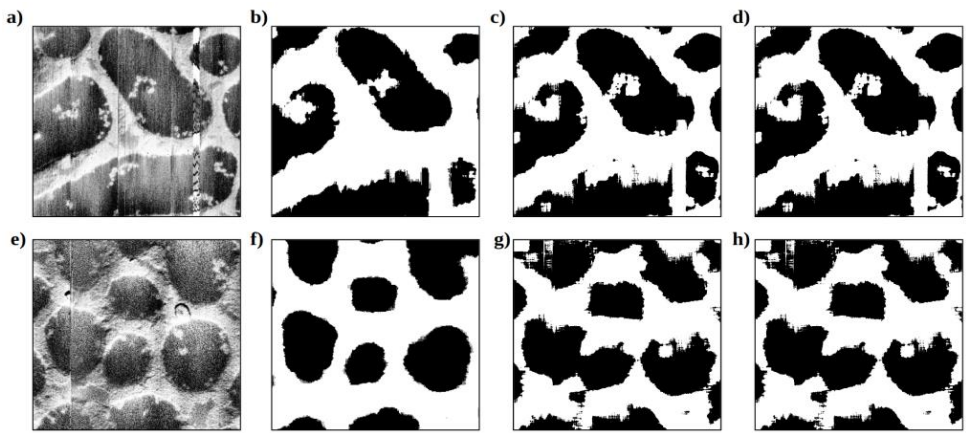

*Figure S9:* To address the input tile size constraint discussed in the main manuscript's **Section II. B**, one could increase the overall image size with image interpolation techniques which can increase the scales of features in tiles. In parts (a) and (e) we show two histogram equalized pre-processed AFM phase images that are sent into the ResNet50 workflow yielding the domain segmented images in parts (b-d) and parts (f-h) when the pre-processed images were interpolated to sizes (c, g) 800 pixels x 800 pixels, (b, f) kept same as input 384 pixels x 384 pixels, and (d, h) 1200 pixels x 1200 pixels. We notice that interpolation has increased the workflow's ability to capture highly granular features (light domains inside larger dark domains are captured in parts c and d as compared to part b. We note, however, that the interpolation method's computational cost scales exponentially with increase in interpolation size.

## 3. Understanding the use of ResNet 50 architecture to extract features

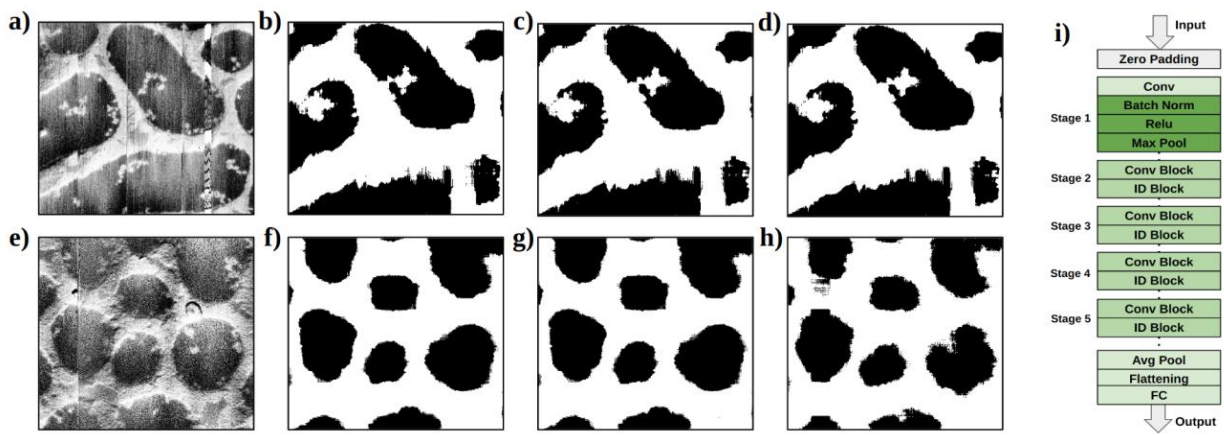

*Figure S10:* ResNet50 is a deep learning algorithm with multiple layers. For the input histogram equalized preprocessed AFM phase images shown in parts (a) and (e), we show the domain segmentation images generated by using feature maps from various stages of ResNet50 model shown in part (i). In parts (b) and (f) are the results for the two inputs when we used feature maps from stage 1 of the flow in part (i). In parts (c) and (g) are results using stage 2 feature maps and in parts (d) and (h) are results using stage3 features maps. With increase in depth, we notice that the workflow becomes more prone to noise [e.g., you can see noise in part (h) that we do not see in parts (f) and (g)]; there can also be misclassification in boundary regions of domains.

## S.VI. Domain Size Distribution Calculated Using Porespy Package

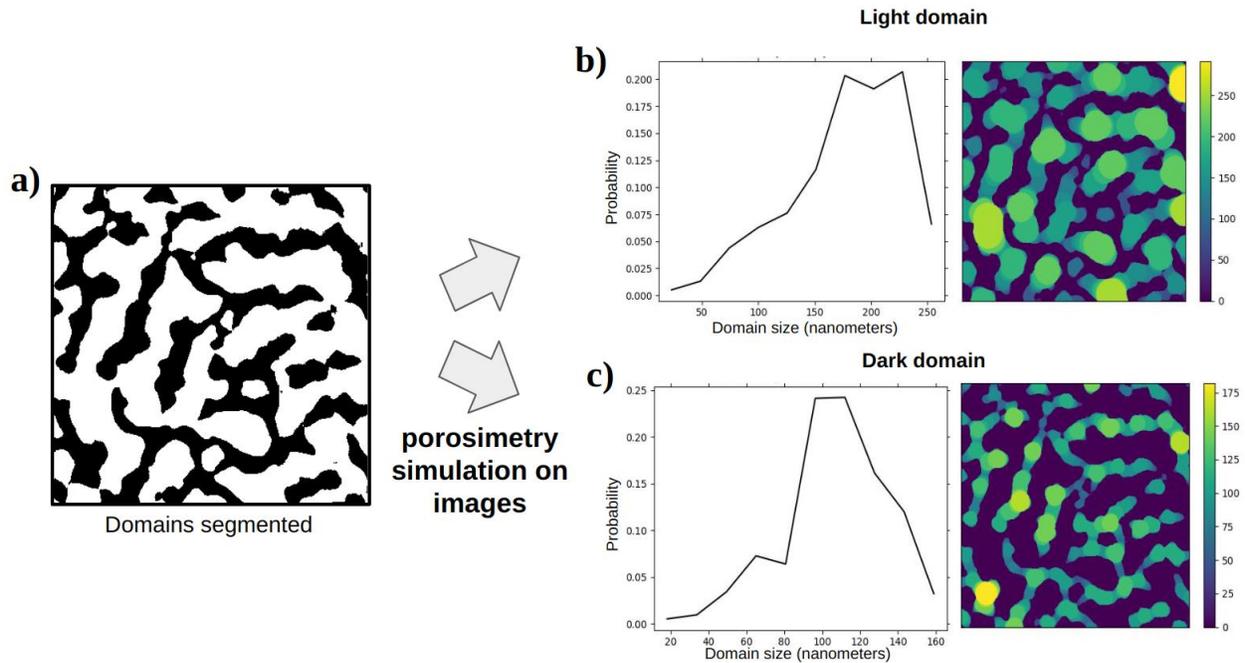

***Figure S11:*** *Illustration of domain size distribution calculations. (a) Index maps (binary images) containing domain segmented into light and dark domains are used to perform domain size distribution calculation.* ***Porespy [4] is a python package*** *that performs porosimetry simulations on index maps resulting in 2D heat maps as shown in the right side of parts (b) and (c). In these heat maps the color of each pixel depicts the radius of the largest circle that could overlap that pixel and the pixels of the non-observing domain are zero. From the heat maps pixel values, we can then calculate domain size distribution in real units using the scale bar present in the metadata associated with the AFM image used in the workflow; parts (b) and (c) on the left present these distributions.*

In **Figure S12** we present the calculated domain size distributions for representative 15 AFM images from the 144 images we had in the dataset. All of these 15 AFM images were segmented using DFT with variance as the features. As described before, these AFM images were obtained from supramolecular block copolymer with varying PS and POEGMA block lengths.

| Pre-processed image | Light domain | Dark domain |

*a).*

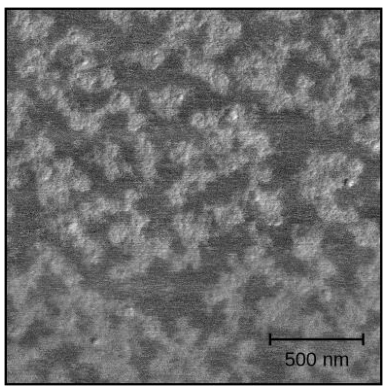 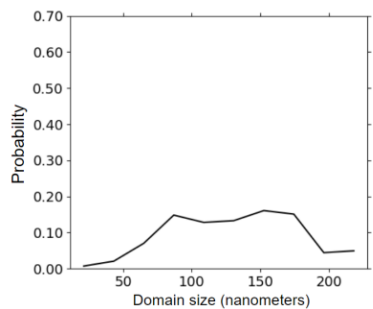 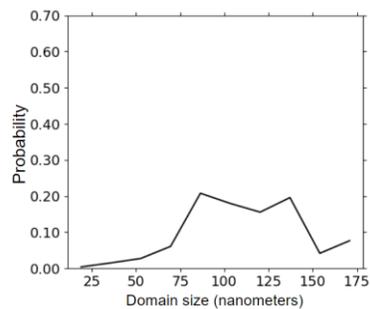

*b).*

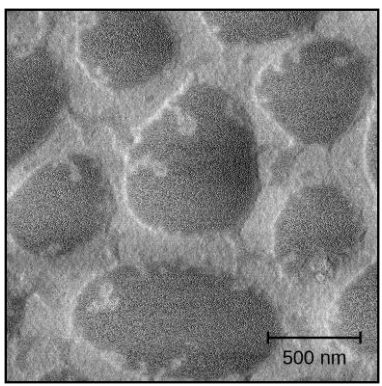 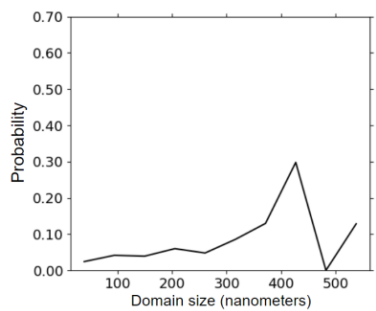 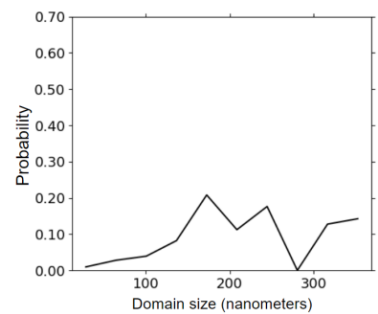

*c).*

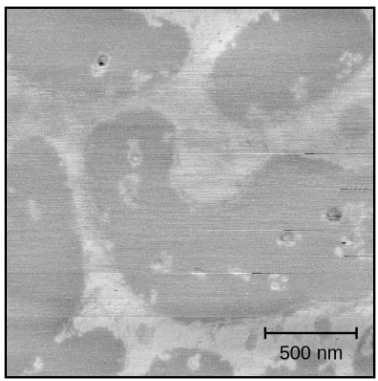 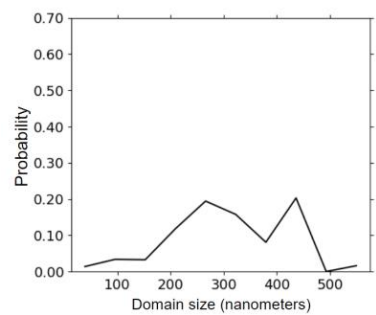 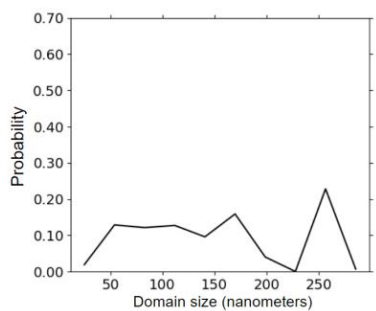

*d).*

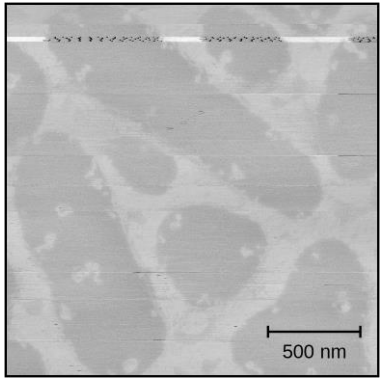 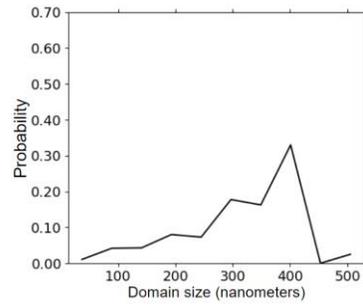 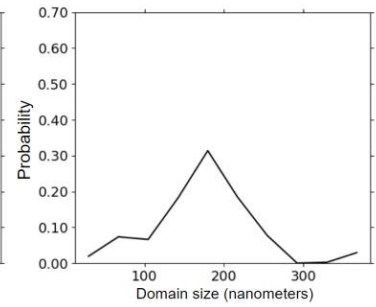

*e).*

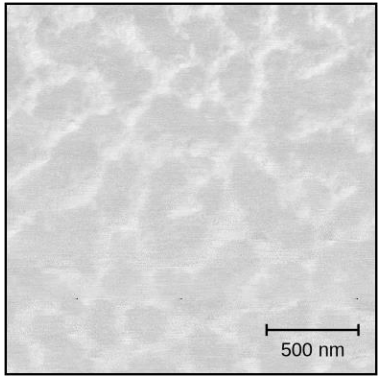 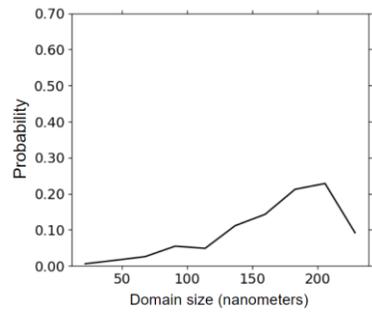 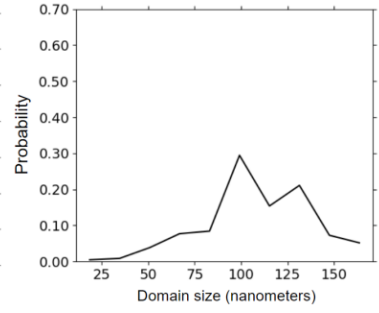

*f).*

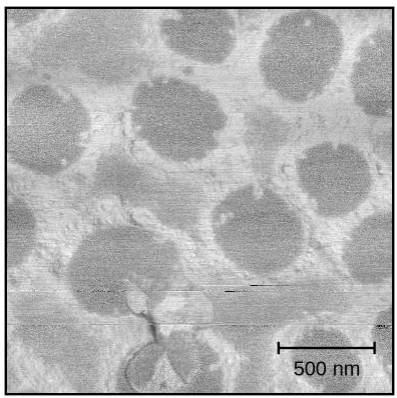 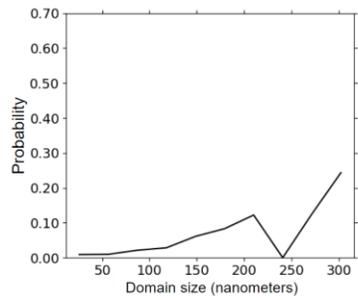 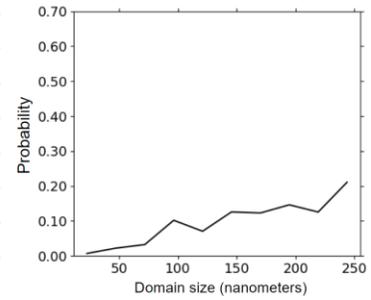

*g).*

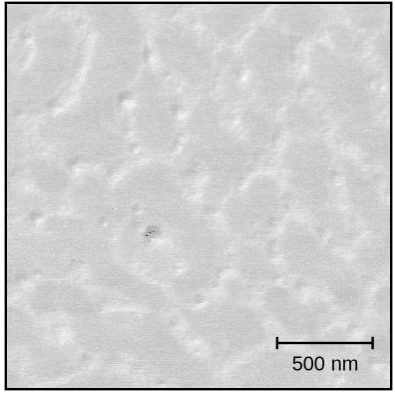 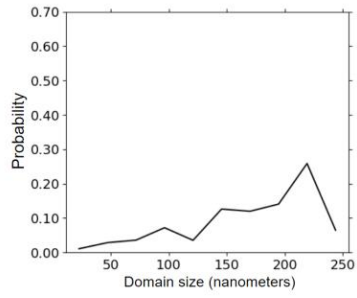 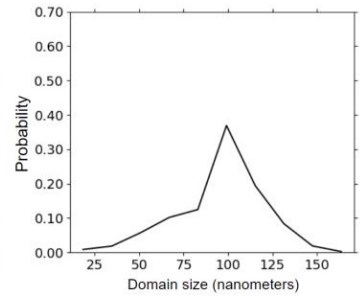

*h).*

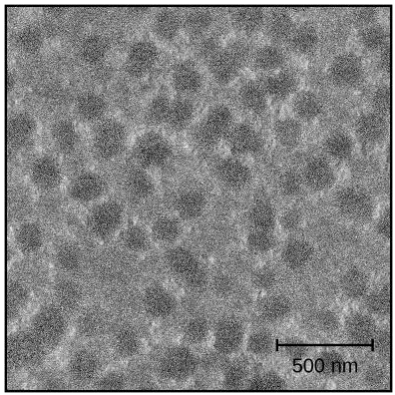 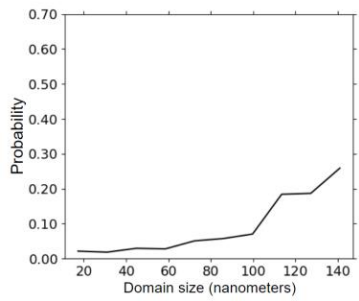 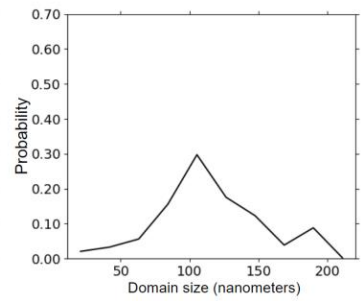

*i).*

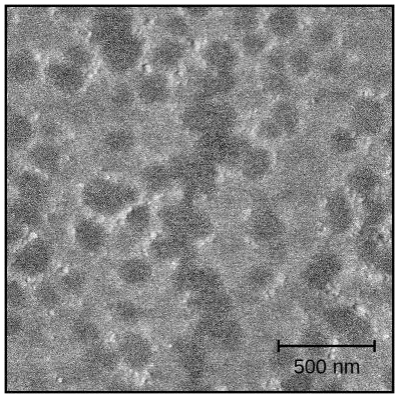 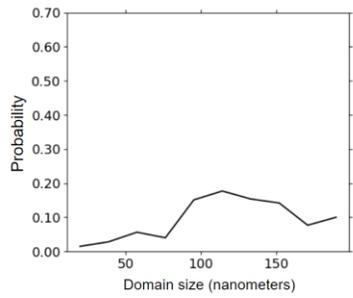 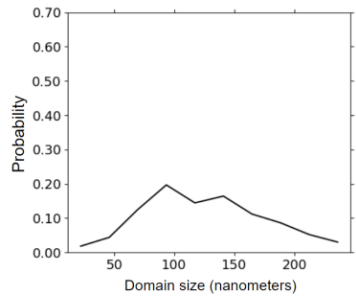

*j).*

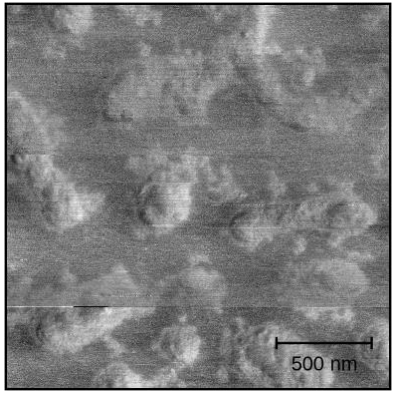 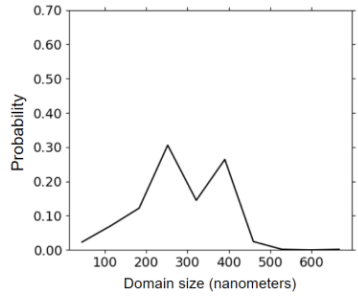 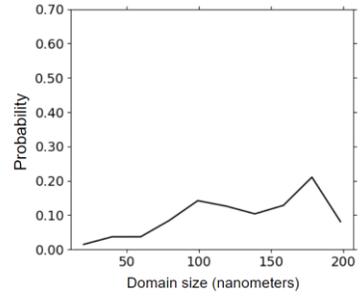

*k).*

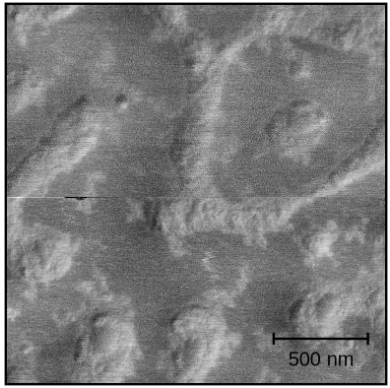 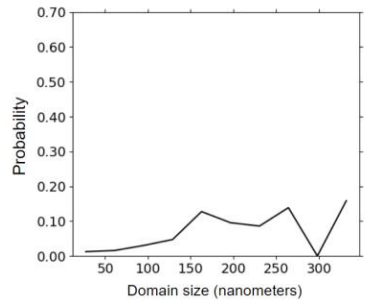 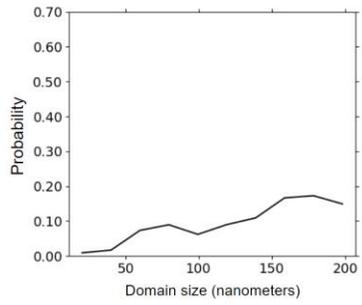

*l).*

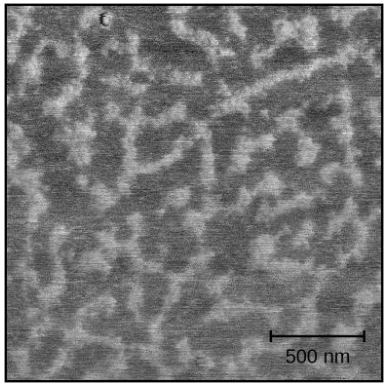 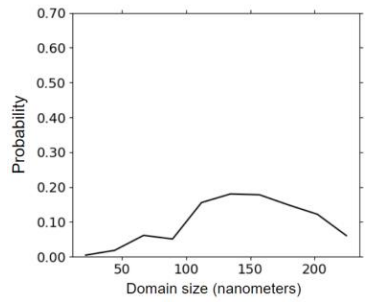 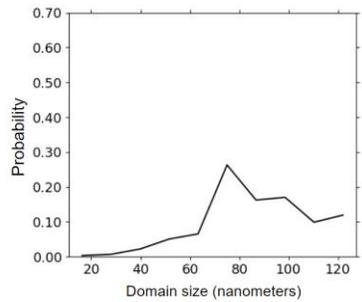

*m).*

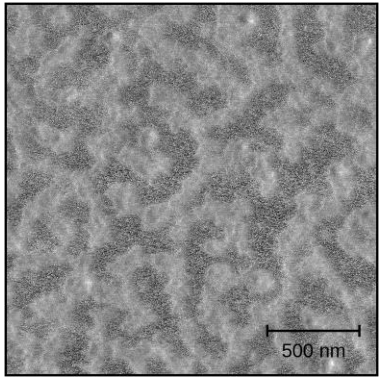 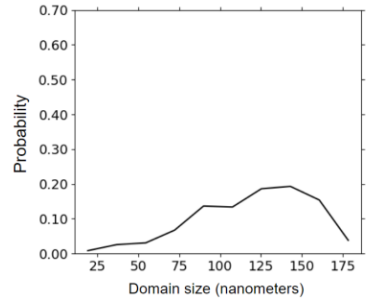 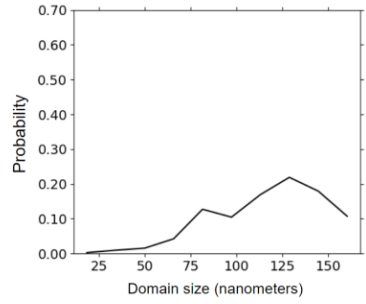

*n).*

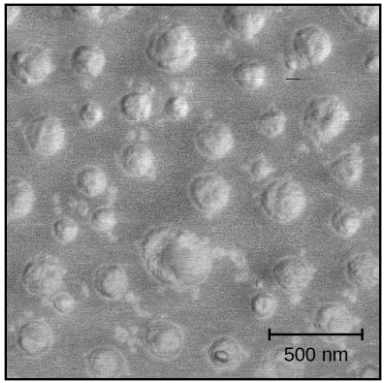 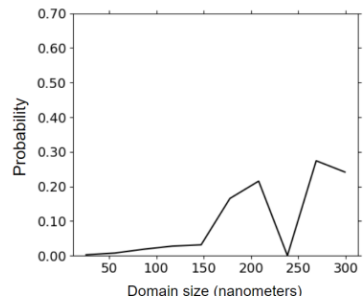 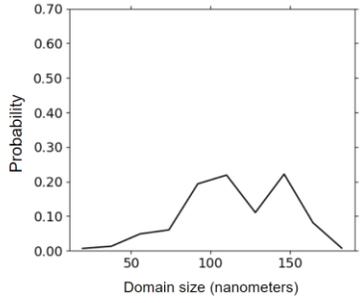

*o).*

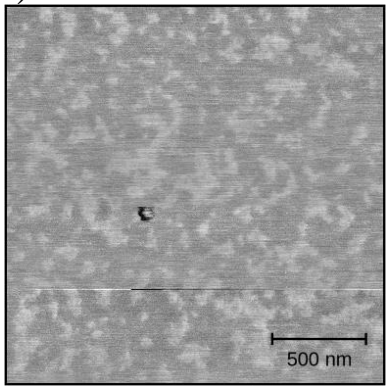 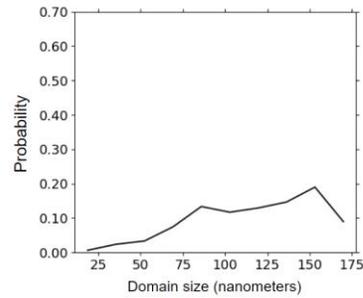 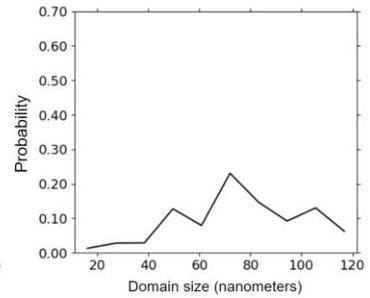

*Figure S12. **Results of domain size distribution for another 15 representative AFM images**. Each panel we have two figures – left is the original AFM pre-processed image and on the right is the domain size distributions (Probability vs. domain sizes in nm) of dark and light domains.*

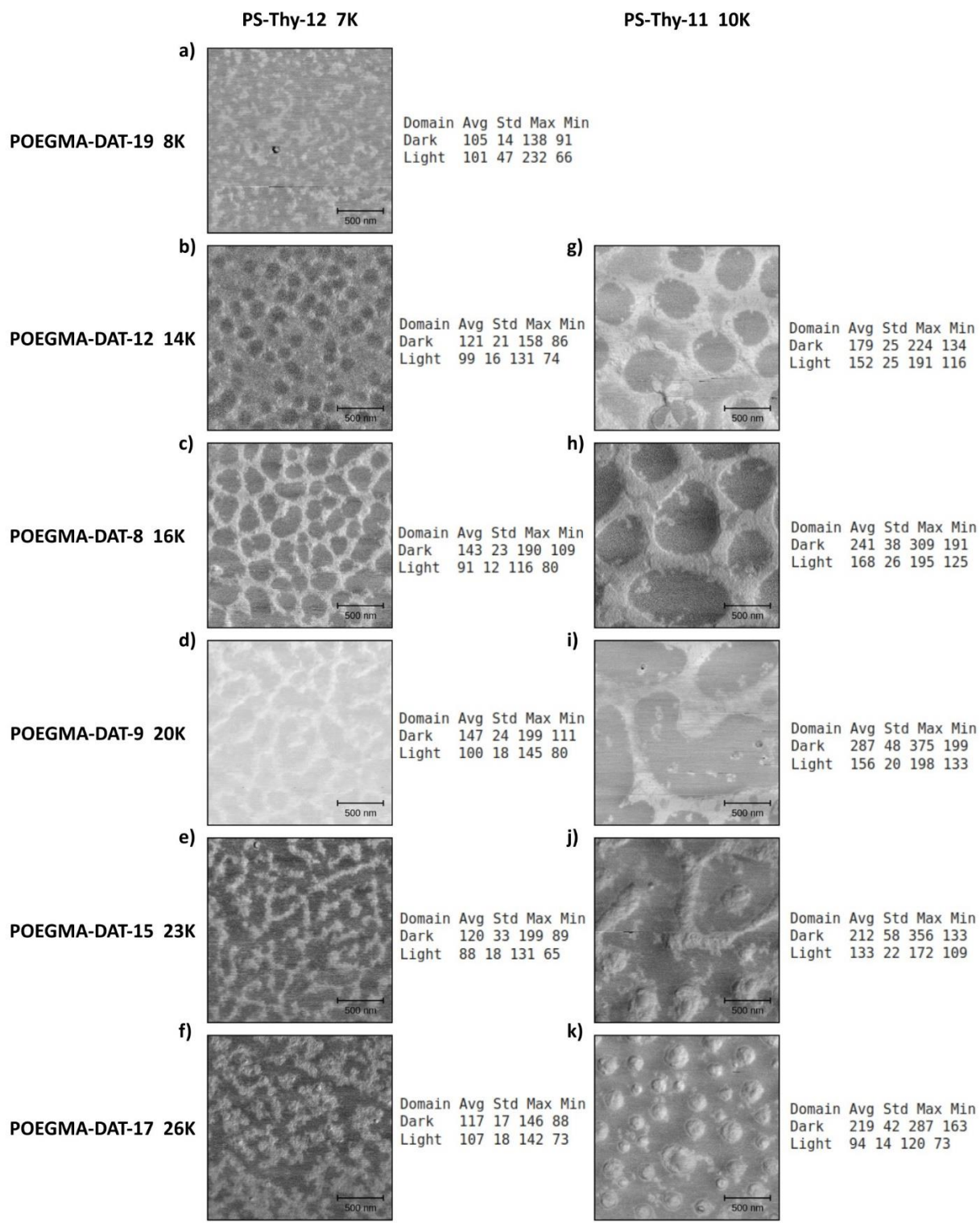

*Figure S13*. Summary of results from domain size distributions for various supramolecular polymers in the AFM dataset shown in *Figure S12*. Here we show the value of the molecular weights of the POEGMA-DAT and PS-Thy in the leftmost column and top-most row. In each part

S.14

*on the left we show the representative AFM image for that system. On the right we share the average, standard deviation, maximum, and minimum of the domain sizes seen for various images collected for each sample with the corresponding molecular weights of the POEGMA-DAT and PS-Thy.*